\documentclass[letterpaper,titlepage,12pt]{article}
\usepackage[colorlinks=true,urlcolor=blue,citecolor=magenta]{hyperref}
\usepackage{amssymb,amsmath,amsfonts}
\usepackage{epsfig}
\usepackage{epsfig}
\usepackage{graphicx}
\usepackage{epstopdf}
\usepackage{caption}
\usepackage{subcaption}
\usepackage{amscd}
\usepackage{amsthm}
\usepackage{latexsym}
\usepackage{amsbsy}
\usepackage{bbm}
\usepackage[english]{babel}
\usepackage{psfrag}
\usepackage{tabularx}

\allowdisplaybreaks

\def\clock{{\count0=\time
           \divide\count0 60
           \ifnum\count0<10 0\fi\the\count0
           \multiply\count0 -60 \advance\count0 \time
           :\ifnum\count0<10 0\fi \the\count0
         }}
\newcommand{\timestamp}{{\small\vbox{\hbox{\tt\jobname.tex}
\hbox{\the\day/\the\month/\the\year, \clock}}}}

\numberwithin{equation}{section}
\begin{document}

\begin{titlepage}
\rightline{\vbox{   \phantom{ghost} }}
%
%\rightline{\vbox{\hfill  CCTP-2015-05
%}}
%\rightline{\vbox{\hfill  CCQCN-2015-64}}
%
 \vskip 1.4 cm
\centerline{\LARGE \bf Ho\v rava--Lifshitz Gravity From Dynamical}
\vspace{.2cm}
\centerline{\LARGE \bf Newton--Cartan Geometry}

\vskip 1.5cm

\centerline{\large {{\bf Jelle Hartong$^1$, Niels A. Obers$^2$}}}

\vskip .8cm

\begin{center}

\sl $^1$ Physique Th\'eorique et Math\'ematique and International Solvay Institutes,\\
Universit\'e Libre de Bruxelles, C.P. 231, 1050 Brussels, Belgium.\\
\sl $^2$ The Niels Bohr Institute, Copenhagen University,\\
\sl  Blegdamsvej 17, DK-2100 Copenhagen \O , Denmark.
\vskip 0.4cm

\end{center}
\vskip 0.6cm

%\centerline{\small\tt  hartong@nbi.dk, obers@nbi.dk}

\vskip .8cm \centerline{\bf Abstract} \vskip 0.2cm \noindent

Recently it has been established that torsional Newton--Cartan (TNC) geometry is the appropriate geometrical framework to which  non-relativistic field theories couple.  We show that when these geometries are made dynamical they give rise to Ho\v rava--Lifshitz (HL) gravity. Projectable HL gravity corresponds to dynamical Newton--Cartan (NC) geometry without torsion and non-projectable HL gravity corresponds to dynamical NC geometry with twistless torsion (hypersurface orthogonal foliation). We build a precise dictionary relating all fields (including the scalar khronon), their transformations and other properties in both HL gravity and dynamical TNC geometry. We use TNC invariance to construct the effective action for dynamical twistless torsional Newton--Cartan geometries in 2+1 dimensions for dynamical exponent $1<z\le 2$ and demonstrate that this exactly agrees with the most general forms of the HL actions constructed in the literature. Further, we identify the origin of the $U(1)$ symmetry observed by Ho\v rava and Melby-Thompson as coming from the Bargmann extension of the local Galilean algebra that acts on the tangent space to TNC geometries. We argue that TNC geometry, which is manifestly diffeomorphism covariant, is a natural geometrical framework underlying HL gravity and  discuss some of its implications.

\end{titlepage}

\pagestyle{empty}
\small
%\begin{spacing}{1}
%\tableofcontents
%\end{spacing}
\tableofcontents

\normalsize
\newpage
\pagestyle{plain}
\setcounter{page}{1}

%\begingroup
%\hypersetup{linkcolor=black}
%\tableofcontents
%\endgroup

\section{Introduction}

In the search for consistent theories of quantum gravity, Ho\v rava-Lifshitz (HL) gravity \cite{Horava:2008ih,Horava:2009uw} has appeared as a tantalizing possibility of a non-Lorentz invariant and renormalizable UV completion of gravity. While observational constraints and  the matching to general relativity in the IR put severe limitations on the phenomenological viability of this proposal, HL gravity is of intrinsic theoretical interest as an example of gravity with anisotropic scaling between time and space. In particular, in the context of holography it holds the prospect of providing an alternative way \cite{Janiszewski:2012nb,Griffin:2012qx} of constructing gravity duals for strongly coupled systems with non-relativistic scaling, including those of interest to condensed matter physics. More generally, one might expect that HL gravity has a natural embedding in the larger framework of string theory \cite{Janiszewski:2012nf}.
 
In parallel to this development, and with in part similar motivations, there has been considerable effort to extend the original AdS-setup in (conventional) relativistic gravity to space-times with non-relativistic scaling \cite{Son:2008ye,Balasubramanian:2008dm,Kachru:2008yh,Taylor:2008tg}. Such space-times typically exhibit a dynamical exponent $z$ that characterizes the anisotropy between time and space on the boundary. This includes in particular holography for Lifshitz space-times, for which it was found that the 
boundary geometry  is described by a novel extension of Newton--Cartan (NC) geometry%
\footnote{We refer to \cite{Dautcourt,Eisenhart,Trautman,Kuenzle:1972zw,Duval:1984cj,Duval:1990hj,Julia:1994bs} for earlier work on Newton--Cartan geometry.}
with a specific torsion tensor, called torsional Newton--Cartan (TNC) geometry. 
The aim of this paper is to construct the theory of dynamical TNC geometry  and show that it exactly agrees with
the most general forms of HL gravity.

TNC geometry was first observed in \cite{Christensen:2013lma,Christensen:2013rfa} as the boundary geometry for a specific action supporting $z=2$ Lifshitz geometries, and subsequently generalized to a large class of holographic Lifshitz models for arbitrary values of $z$ in \cite{Hartong:2014oma,Hartong:2014}. In parallel, it was shown in detail in \cite{Bergshoeff:2014uea} how TNC geometry arises by gauging the Schr\"odinger algebra, following the earlier work  \cite{Andringa:2010it} on obtaining NC geometry from gauging the Bargmann algebra. In this paper we will show that TNC geometry can also be obtained by generalizing directly the work of \cite{Andringa:2010it} to include torsion without using the Schr\"odinger algebra. In its broadest sense the results of \cite{Hartong:2014oma,Hartong:2014} imply that Lifshitz holography describes a dual version of field theories on TNC backgrounds. In \cite{Hartong:2015wxa} it was shown that the Lifshitz vacuum (in Poincar\'e type coordinates) exhibits the same symmetry properties as a flat NC space-time. In particular it was found that the conformal Killing vectors of flat NC space-time span the Lifshitz algebra. In order to understand the properties of field theories on TNC backgrounds some simple scale invariant scalar field models on flat NC space-time were studied in \cite{Hartong:2014pma,Hartong:2015wxa}. It was shown that two scenarios can occur: i). either the theory has an internal local $U(1)$ symmetry related to particle number or ii). it does not. In case i). there is a mechanism that enhances the global Lifshitz symmetries to include particle number and Galilean boosts (and possibly even special conformal transformations) whereas in the other case no such symmetry enhancement can take place. This means that the notion of global symmetries depends on the type of matter fields one considers on such a background. In support of  this it was demonstrated in Ref.~\cite{Hartong:2015wxa} that one can define probe scalars on a Lifshitz background that have a global Schr\"odinger invariance. The field-theoretic perspective of coupling Galilean invariant field theories to TNC%
\footnote{Ref.~ \cite{Son:2013rqa}. introduced NC geometry to field theory analyses of problems with strongly correlated electrons, such as the fractional quantum Hall effect. Later torsion was added to this analysis in \cite{Geracie:2014nka}. The type of torsion introduced there is what we call twistless torsion.
See also \cite{Banerjee:2014nja,Brauner:2014jaa} for a different approach to Newton--Cartan geometry.}
was independently considered in \cite{Jensen:2014aia}.

The relevant geometric fields in TNC are a time-like vielbein $\tau_\mu$, an inverse spatial metric  $h^{\mu \nu}$ and a vector field $M_\mu=m_\mu-\partial_\mu\chi$ where $\chi$ is a St\"uckelberg scalar whose role in TNC geometry will be elucidated in section \ref{sec:torsion}. The torsion in TNC geometry is always proportional to $\partial_\mu\tau_\nu-\partial_\nu\tau_\mu$ where $\tau_\mu$ defines the local flow of time. The amount of torsion depends on the properties of $\tau_\mu$ and we distinguish the three cases:%
\footnote{These three cases also naturally arise in Lifshitz holography  \cite{Christensen:2013lma,Christensen:2013rfa}. We note that  TTNC geometry was already observed in \cite{Julia:1994bs} but in that work the torsion was eliminated using a conformal rescaling.} 
\begin{itemize}
\item Newton--Cartan (NC) geometry
\item twistless torsional (TTNC) geometry
\item torsional Newton--Cartan (TNC) geometry
\end{itemize}
where the first possibility has no torsion and the latter option has general torsion with the twistless case being an important in-between situation. More specifically, in the first case the time-like vielbein of the geometry is closed and defines an absolute time. In the second case the time-like vielbein is hypersurface orthogonal and thereby allows for a foliation of equal time spatial surfaces described by Riemannian (i.e. torsion free) geometry. In the third, most general, case there is no constraint on $\tau_\mu$.

As is clear from holographic studies of the boundary energy-momentum tensor as for example in \cite{Ross:2011gu,Christensen:2013lma,Christensen:2013rfa,Hartong:2014oma,Hartong:2015wxa} the addition of torsion to the NC geometry is crucial in order to be able to calculate the energy density and energy flux of the theory. This is because they are the response to varying $\tau_\mu$ (see also \cite{Jensen:2014aia}). Hence in order to be able to compute these quantities $\tau_\mu$ better be unconstrained, i.e. one should allow for arbitrary torsion. If we work with TTNC geometry one can only compute the energy density and the divergence of the energy current \cite{Hartong:2014} because in that case $\tau_\mu=\psi\partial_\mu\tau$ where one has to vary $\psi$ and $\tau$ with $\psi$ sourcing the energy density and $\tau$ sourcing the divergence (after partial integration) of the energy current. In any case the point is that, contrary to the relativistic setting, adding torsion is a very natural thing to do in NC geometry. Moreover, as will be shown later, the torsion is not something one can freely pick and  is actually fixed by the formalism.

In all of these works the TNC geometry appears as a fixed background and is hence not dynamical. The purpose of this paper is to consider what theory of gravity appears when letting the TNC geometry fluctuate.  We find, perhaps not entirely unexpected%
\footnote{A HL-type action in TNC covariant form was already observed in  \cite{Christensen:2013rfa}
where the anisotropic Weyl-anomaly in a specific $z=2$ holographic four-dimensional bulk Lifshitz model was obtained
via null Scherk--Schwarz reduction of the AdS$_5$ conformal anomaly of gravity coupled to an axion.}, 
that depending on the amount of torsion the resulting theories include  HL gravity and all of its known extensions.

Our focus in this paper will be mainly on the first two of the three cases listed above, leaving the details of the dynamics of the most general case (TNC gravity) for future work. 
In particular, we will show that: 
\begin{itemize}
\item dynamical NC geometry = projectable HL gravity
\item dynamical  TTNC geometry = non-projectable HL gravity.
\end{itemize}
The khronon field introduced by \cite{Blas:2010hb} (to make HL gravity generally covariant whereby making manifest the presence of an extra scalar mode) naturally appears (see also \cite{Germani:2009yt}) in our formulation. We furthermore show  that the $U(1)$ extension of  \cite{Horava:2010zj} (see also \cite{daSilva:2010bm,Zhu:2011yu}) emerges as well in a natural fashion. The essential identification between the covariant%
\footnote{Note that in e.g. Ref.~\cite{Jacobson:2010mx} there is also a type of covariantization of HL gravity (see also eq. (3.9) of \cite{Janiszewski:2012nb}), but there is still inherently a Lorentzian metric structure present. This only works up to second order in derivatives so that it only captures the IR limit of HL gravity.}
NC-type geometric structures and those appearing in the ADM parametrization that forms the starting point of HL gravity is as follows
$$
\tau_\mu \sim {\rm lapse} \quad, \quad  \hat h_{\mu\nu}  \sim \mbox{spatial metric}   \quad , \quad
m_\mu \sim \mbox{shift + Newtonian potential} \,,
$$
where the fields $\hat h_{\mu\nu}$ and $m_\mu$ are defined in section \ref{sec:Bargmann}. We will show that the effective action for the TTNC fields leads to two kinetic terms for the metric $\hat h_{\mu\nu}$ (giving rise to the $\lambda$ parameter of HL gravity \cite{Horava:2008ih,Horava:2009uw}) including the potential terms computed in Refs.~\cite{Blas:2009qj,Blas:2010hb,Zhu:2011yu}.

Furthermore the St\"uckelberg scalar $\chi$ entering in the TNC quantity $M_\mu = m_\mu - \partial_\mu \chi$ (see \cite{Christensen:2013lma,Christensen:2013rfa,Hartong:2014oma,Hartong:2014pma,Bergshoeff:2014uea,Hartong:2015wxa}) will be directly related to the Newtonian prepotential introduced in \cite{Horava:2010zj}. The relation to TTNC geometry will, however, provide a new perspective on the nature of the $U(1)$ symmetry studied in the context of HL gravity. As a further confirmation that TNC geometry is a natural framework for HL gravity we will demonstrate in this paper that when we include dilatation symmetry (local Schr\"odinger invariance) one obtains conformal HL gravity.
 
As we will review in this paper, the various versions of TNC geometry defined above arise by gauging non-relativistic symmetry algebras (Galilean, Bargmann, Schr\"odinger). In particular, in this procedure the internal symmetries are made into local symmetries, and translations are turned into diffeomorphisms. This is in the same way that Riemannian geometry comes from gauging the Poincar\'e algebra, thereby imposing local Lorentz symmetry and turning translations into space-time diffeomorphisms. Thus HL gravity theories (and more generally TNC gravity) can be seen as the most general gravity theories for which the Einstein equivalence principle (that locally space-time is described by flat Minkowski space-time) is applied to {\it local} non-relativistic (Galilean) symmetries, rather than to the local Lorentz symmetry that one has in special relativity.

We point out that in general relativity (GR) the global symmetries (Killing vectors) of Minkowski space-time (the Poincar\'e algebra) form the same algebra from which upon gauging (and replacing local space-time translations by diffeomorphisms as explained in appendix \ref{app:gaugingPoincare}) we obtain the geometrical framework of GR. On the other hand the Killing vectors of flat NC space-time only involve space and time translations and spatial rotations \cite{Hartong:2015wxa} while the local tangent space group that we gauge in order to obtain the TNC geometrical framework is the Galilean algebra (where again we also replace local time and space translations by diffeomorphisms), which also contains Galilean boosts and is thus not the same algebra as the algebra of Killing vectors of flat NC space-time. We bring this up to highlight the fact that the local tangent space symmetries and the Killing vectors of flat space-time are in general two very different concepts that are often mistakenly assumed to be the same. Basically this happens because the $M_\mu$ vector allows for the construction of a new set of vielbeins (defined in section \ref{sec:Bargmann}) that are invariant under $G$ transformations and that only see diffeomorphisms and local rotations which agrees with the Killing vectors of flat NC space-time. Nevertheless the fact that $M_\mu$ is one of the background fields to which we can couple a field theory can, under special circumstances, lead to additional symmetries such as $G$ and $N$ (and even special conformal symmetries) \cite{Hartong:2015wxa}.

Our results on dynamical TNC geometry and its relation to HL gravity provide a new perspective on these theories of gravity. For one thing, the vacuum of HL gravity (without a cosmological constant) has so far  been taken to be Minkowski space-time, but since the underlying geometry appears to be TNC geometry, it seems more natural to take this as flat NC space-time  \cite{Hartong:2014pma,Hartong:2015wxa}. Thus it would seem worthwhile to reexamine HL gravity and the various issues%
\footnote{There is an extensive literature on this (e.g. instabilities and strong coupling  at low energies), see e.g. Refs.  \cite{Cai:2009dx,Charmousis:2009tc,Li:2009bg,Sotiriou:2009bx,Blas:2009yd,Bogdanos:2009uj,Koyama:2009hc,Henneaux:2009zb}.}
that have been raised following its introduction. 
As another application, we emphasize that, independent of a possible UV completion of gravity, our results on dynamical TNC geometry are of relevance to constructing IR effective field theories of non-relativistic systems following the recent developments of applying this to condensed matter systems.  
For these kinds of applications, the question whether HL gravity flows to a theory with
local Lorentz invariance ($\lambda=1$) in the IR is of no concern.   Finally, from a broader perspective
our results might be useful  towards a proper description of the non-relativistic quantum gravity corner of the 
``$(\hbar,G_N,1/c)$-cube'',  perhaps aiding the formulation of a well-defined  perturbative $1/c$ expansion around such a theory.

\subsubsection*{Outline of the paper}

The first part of the paper (sections \ref{sec:LocalGalTrafos} to \ref{sec:curvatures}) is devoted to setting up the geometrical framework for torsional Newton--Cartan geometry, presented in such a way that the subsequent connection to HL gravity is most clearly displayed. We thus take a pedagogical approach that introduces the relevant ingredients in a step-by-step way. To this end we begin in section  \ref{sec:LocalGalTrafos} with the geometry that is obtained by gauging the Galilean algebra, extending the original work of \cite{Andringa:2010it} to include torsion. We exhibit the transformation properties of the relevant geometrical fields under space-time diffeomorphisms and the internal transformations, consisting of Galilean boosts ($G$) and spatial rotations ($J$). We also discuss the vielbein postulates and curvatures entering the field strength of the gauge field.  We point out that the only $G,J$ invariants are the time-like vielbein $\tau_\mu$ and the inverse spatial metric $h^{\mu \nu}$. In section 
\ref{sec:AC1} we then present the most general affine connection that satisfies the property that the latter quantities are covariantly conserved. 

In section \ref{sec:Bargmann}, we go one step further and add the central element ($N)$ to the Galilean algebra, and consider the gauging of the resulting Bargmann algebra (as also considered in \cite{Andringa:2010it} for the case with no torsion). We show that the extra gauge field $m_\mu$ that enters in this description, does not alter the transformation properties of the objects considered in section \ref{sec:LocalGalTrafos}, but  allows for the introduction of further useful $G,J,$ invariants, namely an inverse time-like vielbein  $\hat v^\mu$, a spatial metric $\bar h_{\mu \nu}$ (or $\hat h_{\mu \nu}$) and a ``Newtonian potential'' $\tilde \Phi$. We then return to the construction of the affine connection in section \ref{sec:connection2}  and employ the geometric quantities of section \ref{sec:LocalGalTrafos} and \ref{sec:Bargmann} to construct the most general connection that can be built out of the invariants. We discuss two special choices of affine connections with particular properties, one of them being especially convenient for the comparison with HL gravity. We point out that, in the case of non-vanishing torsion, there is no choice of affine connection that is also $N$-invariant, but that one can formally remedy this by introducing a St\"uckelberg scalar $\chi$ (defining $M_\mu = \mu_\mu - \partial_\mu \chi$) to the setup that cancels this non-invariance. This has the advantage that one can deal simultaneously with theories that have a local $U(1)$ symmetry and those that do not have this, and further it will prove useful when comparing to HL gravity (especially  \cite{Horava:2010zj,daSilva:2010bm,Zhu:2011yu}). We also show how the TNC invariants can be used to build a non-degenerate symmetric rank 2 tensor with Lorentzian signature, which will later be used to make contact with the ADM decomposition that enters HL gravity. 

In section \ref{sec:torsion} we discuss the specific form of the torsion tensor that emerges from gauging the Bargmann algebra and introduce the three relevant cases for torsion (NC, TTNC and TNC) that were already mentioned above. We also introduce a vector $a_\mu$ that describes the TTNC torsion, which will turn out to be very useful in order to make contact with the literature on non-projectable HL gravity. Further we will identify the khronon field of \cite{Blas:2010hb}. Then in section \ref{sec:curvatures}  we give some basic properties of the curvatures (extrinsic curvature and Ricci tensor for TTNC) that will be useful when constructing HL actions. 

In section \ref{sec:parametrization} we relate the TNC invariants  introduced in the previous sections to those appearing in the corresponding ADM parameterization employed in HL gravity. This identification and the match of the properties and number of components and local symmetries in the case of NC and TTNC already strongly suggest that dynamical (TT)NC is expected to be the same as (non)-projectable HL gravity. We then proceed in section \ref{sec:HLactions} by showing that the generic action that describes dynamical TTNC geometries agrees on the nose with the most general HL actions appearing in the literature.  For simplicity we treat the case of 2 spatial dimensions with $1<z\le 2$ and organize the terms in the action according to their dilatation weight. In particular, we construct all $G,J$ invariant terms that are relevant or marginal, using as building blocks the TNC invariants (including the torsion tensor and curvature tensor) and covariant derivatives. The resulting action is written in \eqref{eq:HLaction}, \eqref{eq:potential} and gives the HL
kinetic terms \cite{Horava:2008ih,Horava:2009uw} while the potential is exactly the same as the 3D version of the potential given in \cite{Blas:2009qj,Blas:2010hb,Zhu:2011yu}. 

We then proceed in section \ref{sec:localU1}  to consider the extension of the action to include invariance under the central extension $N$, leading to HL actions with local Bargmann invariance. 
This can be achieved by including couplings to $\tilde \Phi$,  which did not appear yet in section 
\ref{sec:HLactions}. Importantly, in the projectable case with the HL coupling constant $\lambda=1$ we reproduce the $U(1)$ invariant action of \cite{Horava:2010zj}. When we consider the non-projectable version or $\lambda \neq 1$ we need additional terms to make the theory $U(1)$ invariant which is precisely achieved by adding the St\"uckelberg field $\chi$ that we introduced in section  \ref{sec:connection2}  (see also \cite{Hartong:2015wxa}). We can then write a Bargmann invariant action that precisely reproduces the actions considered in the literature, where in particular the $\chi$-dependent pieces agree with those in \cite{Zhu:2011yu}. 
This comes about in part via coupling to the natural TNC Newton potential, $\tilde \Phi_\chi$, which is the Bargmann invariant generalization of $\tilde \Phi$, and the simple covariant form of the action \eqref{eq:HLaction4} is one of our central results. 

We emphasize that adding the $\chi$ field to the action means that we have trivialized the $U(1)$ symmetry by St\"uckelberging it or in other words we have removed the $U(1)$ transformations all together. We further expand on this fact in section \ref{sec:constraint}, commenting on statements in the literature regarding the relevance of the $U(1)$ invariance (which is not there unless we have zero torsion and $\lambda=1$) in relation to the elimination of a scalar degree of freedom. In particular, we will present a different mechanism that accomplishes this and which involves a constraint equation obtained by varying the TNC potential $\tilde \Phi_\chi$. 

Finally in section \ref{sec:confHLactions}  we consider the case where we add dilatations to the Bargmann algebra, i.e. we consider the dynamics we get from a geometry that is locally Schr\"odinger invariant. We will show that the resulting theory is conformal HL gravity, providing further evidence for our claim that TNC geometry is the underlying geometry of HL gravity. In particular, employing the local Schr\"odinger algebra we will arrive at the invariant $z=d$ action \eqref{eq:confinvaction} for conformal HL gravity in $d+1$ dimensions.

We end in section \ref{sec:discussion} with our conclusions and discuss a large variety of possible open directions. For comparison to general relativity and as an introduction to the logic followed in sections 
 \ref{sec:LocalGalTrafos} to \ref{sec:curvatures}, we have included appendix \ref{app:gaugingPoincare} which discusses the gauging of the Poincar\'e algebra leading to Riemannian geometry (possibly with torsion added).

\section{Local Galilean Transformations}\label{sec:LocalGalTrafos}

The present section until section \ref{sec:curvatures} is devoted to setting up the general geometrical framework for torsional Newton--Cartan geometry. We will follow an approach that is very similar to what in general relativity is known as the gauging of the Poincar\'e algebra. This provides us in a very efficient manner with all basic geometrical objects used in the formulation of general relativity (and higher curvature modifications thereof). For the interested reader unfamiliar with this method we give a short summary of it in appendix \ref{app:gaugingPoincare}. 

To obtain torsional Newton--Cartan geometry we follow the same logic as in appendix \ref{app:gaugingPoincare} for the case of the Galilean algebra and its central extension known as the Bargmann algebra. This was first considered in \cite{Andringa:2010it} for the case without torsion. Here we generalize this interesting work to the case with torsion. Adding torsion to Newton--Cartan geometry can also be done by making it locally scale invariant, i.e. gauging the Schr\"odinger algebra as in \cite{Bergshoeff:2014uea}. However upon gauging the Schr\"odinger algebra the resulting geometric objects are all dilatation covariant which is useful for the construction of conformal HL gravity as we will study in section \ref{sec:confHLactions} but it is less useful for the study of general non-conformally invariant HL actions which is why we start our analysis by adding torsion to the analysis of \cite{Andringa:2010it}.

Consider the Galilean algebra whose generators are denoted by $H, P_a, G_a, J_{ab}$ and whose commutation relations are 
\begin{equation}\label{eq:Galalgebra}
\begin{array}{ll}
\left[H\,,G_a\right] = P_a\,, & \left[P_a\,,G_b\right] = 0\,,\\
\left[J_{ab}\,,P_c\right] = \delta_{ac}P_b-\delta_{bc}P_a\,, &
\left[J_{ab}\,,G_c\right] = \delta_{ac}G_b-\delta_{bc}G_a\,,\\
\left[J_{ab}\,,J_{cd}\right] = \delta_{ac}J_{bd}-\delta_{ad}J_{bc}-\delta_{bc}J_{ad}+\delta_{bd}J_{ac}\,. &
\end{array}
\end{equation}
Let us consider a connection $\mathcal{A}_\mu$ taking values in the Galilean algebra%
\footnote{Our notation is such that $\mu,\nu=0 \ldots d $ are spacetime indices and $a,b = 1 \dots d$ are spatial
tangent space indices.}
\begin{equation}\label{eq:Schconnection}
\mathcal{A}_\mu = H\tau_{\mu}+P_ae_{\mu}^a+G_a\Omega_{\mu}{}^a+\frac{1}{2}J_{ab}\Omega_{\mu}{}^{ab}\,.
\end{equation}
 This connection transforms in the adjoint as
\begin{equation}\label{eq:YMtrafo}
\delta{\mathcal{A}}_{\mu}=\partial_\mu\Lambda+[{\mathcal{A}}_{\mu}\,,\Lambda]\,.
\end{equation}
With this transformation we can associate another transformation denoted by $\bar\delta$ as follows. Write (without loss of generality)
\begin{equation}
\Lambda=\xi^\mu\mathcal{A}_\mu+\Sigma\,,
\end{equation}
where
\begin{equation}
\Sigma = G_a\lambda^a+\frac{1}{2}J_{ab}\lambda^{ab}\,,
\end{equation}
is chosen to only include the internal symmetries $G$ and $J$. We define $\bar\delta\mathcal{A}_\mu$ as
\begin{equation}\label{eq:bardeltaA}
\bar\delta\mathcal{A}_\mu=\delta\mathcal{A}_\mu-\xi^\nu \mathcal{F}_{\mu\nu}=\mathcal{L}_\xi\mathcal{A}_\mu+\partial_\mu\Sigma+[{\mathcal{A}}_{\mu}\,,\Sigma]\,,
\end{equation}
where $\mathcal{F}_{\mu\nu}$ is the curvature
\begin{eqnarray}
\mathcal{F}_{\mu\nu} & = & \partial_\mu\mathcal{A}_\nu-\partial_\nu\mathcal{A}_\mu+[{\mathcal{A}}_{\mu}\,,{\mathcal{A}}_{\nu}]\nonumber\\
&=& HR_{\mu\nu}(H)+P_a R_{\mu\nu}{}^a(P)+G_a R_{\mu\nu}{}^a(G)+\frac{1}{2}J_{ab}R_{\mu\nu}{}^{ab}(J)\,.
\label{FGal}
\end{eqnarray}
Often in works on gauging space-time symmetry groups it is suggested that diffeomorphisms can only be obtained once specific curvature constraints are imposed\footnote{This is because setting to zero some of the curvatures in $\mathcal{F}_{\mu\nu}$ identifies $\bar\delta$ with $\delta$ in \eqref{eq:bardeltaA} for those fields that are not fixed by the curvature constraints. There is no need for the $\delta$ and $\bar\delta$ transformations to coincide. As we show in appendix \ref{app:gaugingPoincare} this is no longer the case in GR when there is non-vanishing torsion.}. We emphasize that the transformation $\bar\delta\mathcal{A}_\mu$ exists no matter what we choose for the curvature $\mathcal{F}_{\mu\nu}$.

If we write in components what \eqref{eq:bardeltaA} states we obtain the transformation properties
\begin{eqnarray}
\bar\delta\tau_\mu & = & \mathcal{L}_\xi\tau_\mu\,,\\
\bar\delta e_\mu^a & = & \mathcal{L}_\xi e_\mu^a+\lambda^a{}_b e^b_\mu+\lambda^a\tau_\mu\,,\label{eq:trafo2}\\
\bar\delta\Omega_\mu{}^a & = & \mathcal{L}_\xi\Omega_\mu{}^a+\partial_\mu\lambda^a+\lambda^a{}_b \Omega_\mu{}^b+\lambda^b\Omega_{\mu b}{}^a\,,\label{eq:trafo3}\\
\bar\delta\Omega_\mu{}^{ab} & = & \mathcal{L}_\xi\Omega_\mu{}^{ab}+\partial_\mu\lambda^{ab}+2\lambda^{[a}{}_c\Omega_\mu{}^{|c|b]}\,,
\end{eqnarray}
where $\mathcal{L}_\xi$ is the Lie derivative along $\xi^\mu$ and $\lambda^a$, $\lambda^{ab}$ the parameters of the internal $G$, $J$ transformations, respectively.

We can now write down covariant derivatives that transform covariantly under these transformations. They are 
\begin{eqnarray}
\mathcal{D}_\mu\tau_\nu & = &\partial_\mu\tau_\nu-\Gamma^\rho_{\mu\nu}\tau_\rho\,,  \label{eq:covdertau}\\
\mathcal{D}_\mu e_\nu^a & = & \partial_\mu e_\nu^a-\Gamma^\rho_{\mu\nu}e_\rho^a-\Omega_\mu{}^a\tau_\nu-\Omega_\mu{}^{a}{}_be_{\nu}^b \,,\label{eq:covdere}
\end{eqnarray}
where $\Gamma^\rho_{\mu\nu}$ is an affine connection transforming as
\begin{equation}\label{eq:affinecon}
\bar\delta \Gamma^\rho_{\mu\nu}=\partial_\mu\partial_\nu\xi^\rho+\xi^\sigma\partial_\sigma\Gamma^\rho_{\mu\nu}+\Gamma^\rho_{\sigma\nu}\partial_\mu\xi^\sigma+\Gamma^\rho_{\mu\sigma}\partial_\nu\xi^\sigma-\Gamma^\sigma_{\mu\nu}\partial_\sigma\xi^\rho\,.
\end{equation}
It is in particular inert under the $G$ and $J$ transformations. The form of the covariant derivatives is completely fixed by the local transformations $\bar\delta\mathcal{A}_\mu$. However any tensor redefinition of the connections $\Gamma^\rho_{\mu\nu}$, $\Omega_\mu{}^a$ and $\Omega_\mu{}^{ab}$ that leaves the covariant derivatives form-invariant leads to an allowed set of connections with the exact same transformation properties. 

We impose the vielbein postulates
\begin{eqnarray}
\mathcal{D}_\mu\tau_\nu & = & 0\,, \label{eq:VP-1} \\
\mathcal{D}_\mu e_\nu^a & = & 0 \,,\label{eq:VP-2}
\end{eqnarray}
which allows us to express $\Gamma^\rho_{\mu\nu}$ in terms of $\Omega_\mu{}^a$ and $\Omega_\mu{}^{ab}$ via
\begin{equation}\label{eq:relationconnections}
\Gamma^\rho_{\mu\nu}=-v^\rho\partial_\mu\tau_\nu+e^\rho_a\left(\partial_\mu e_\nu^a-\Omega_\mu{}^a\tau_\nu-\Omega_\mu{}^{a}{}_be_{\nu}^b\right)\,,
\end{equation}
where we defined inverse vielbeins $v^\mu$ and $e^\mu_a$ via
\begin{equation}
v^\mu\tau_\mu=-1\,,\qquad v^\mu e_\mu^a=0\,,\qquad e^\mu_a\tau_\mu=0\,,\qquad e^\mu_a e_\mu^b=\delta^b_a\,.
\end{equation}
The vielbein postulates for the inverses read
\begin{eqnarray}
\mathcal{D}_\mu v^\nu & = & \partial_\mu v^\nu+\Gamma^\nu_{\mu\rho}v^\rho-\Omega_\mu{}^a e^\nu_a=0\,,\label{eq:invVP1}  \\
\mathcal{D}_\mu e^\nu_a & = & \partial_\mu e^\nu_a+\Gamma^\nu_{\mu\rho}e^\rho_a+\Omega_\mu{}^{b}{}_ae^\nu_b=0 \,.\label{eq:invVP2}
\end{eqnarray}
Using that $\Omega_\mu{}^{ab}$ is antisymmetric we find that 
\begin{equation}\label{eq:conditionGamma1}
\nabla_\mu h^{\nu\rho}=0\,,
\end{equation}
which together with equations \eqref{eq:covdertau} and \eqref{eq:VP-1}, i.e.
\begin{equation}\label{eq:conditionGamma2}
\nabla_\mu\tau_\nu=0\,,
\end{equation}
constrain $\Gamma^\rho_{\mu\nu}$. Equations \eqref{eq:conditionGamma1} and \eqref{eq:conditionGamma2} are the TNC analogue of metric compatibility in GR.

The components of the field strength $\mathcal{F}_{\mu\nu}$ in \eqref{FGal} are given by
\begin{eqnarray}
R_{\mu\nu}(H) & = & 2\partial_{[\mu}\tau_{\nu]}\,,\\
R_{\mu\nu}{}^a(P) & = & 2\partial_{[\mu} e_{\nu]}^a-2\Omega_{[\mu}{}^a\tau_{\nu]}-2\Omega_{[\mu}{}^{a}{}_be_{\nu]}^b\,,\label{eq:RP}\\
R_{\mu\nu}{}^a(G) & = & 2\partial_{[\mu}\Omega_{\nu]}{}^a-2\Omega_{[\mu}{}^{ab}\Omega_{\nu]b}\,,\\
R_{\mu\nu}{}^{ab}(J) & = & 2\partial_{[\mu} \Omega_{\nu]}{}^{ab}-2\Omega_{[\mu}{}^{ca}\Omega_{\nu]}{}^b{}_c\,.
\end{eqnarray}
The first two appear in the antisymmetric part of the covariant derivatives $\mathcal{D}_\mu\tau_\nu$ and $\mathcal{D}_\mu e_\nu^a$. More precisely we have
\begin{eqnarray}
R_{\mu\nu}(H) & = & 2\Gamma^\rho_{[\mu\nu]}\tau_\rho\,,\label{eq:curvH}\\
R_{\mu\nu}{}^a(P) & = & 2\Gamma^\rho_{[\mu\nu]}e_\rho^a\,.\label{eq:curvP}
\end{eqnarray}
In other words they are equal to the torsion tensor, i.e.
\begin{equation}
2\Gamma^\rho_{[\mu\nu]}=-v^\rho R_{\mu\nu}(H)+e^\rho_a R_{\mu\nu}{}^a(P)\,.
\end{equation}
The other two curvature tensors can be found by computing the Riemann tensor defined as
\begin{equation}\label{eq:nablacommutator}
[\nabla_\mu\,,\nabla_\nu ]X_\sigma=R_{\mu\nu\sigma}{}^\rho X_\rho-2\Gamma^\rho_{[\mu\nu]}\nabla_\rho X_\sigma\,.
\end{equation}
Using that 
\begin{equation}\label{eq:Riemann}
R_{\mu\nu\sigma}{}^\rho=-\partial_\mu\Gamma^\rho_{\nu\sigma}+\partial_\nu\Gamma^\rho_{\mu\sigma}-\Gamma^\rho_{\mu\lambda}\Gamma^\lambda_{\nu\sigma}+\Gamma^\rho_{\nu\lambda}\Gamma^\lambda_{\mu\sigma}\,,
\end{equation}
together with \eqref{eq:relationconnections} tells us that
\begin{equation}\label{eq:decompRiemann}
R_{\mu\nu\sigma}{}^\rho=e^{\rho}_a\tau_\sigma R_{\mu\nu}{}^a(G)-e_{\sigma a}e^\rho_b R_{\mu\nu}{}^{ab}(J)\,.
\end{equation}

So far all components of $\mathcal{A}_\mu$ are independent or what is the same $\tau_\mu$, $e_\mu^a$ and $\Gamma^\rho_{\mu\nu}$ (obeying \eqref{eq:conditionGamma1} and \eqref{eq:conditionGamma2}) are all independent. The inverse vielbeins $v^\mu$ and $e^\mu_a$ transform as
\begin{eqnarray}
\bar\delta v^\mu & = & \mathcal{L}_\xi v^\mu+e^\mu_a\lambda^a\,,\\
\bar\delta e^\mu_a & = & \mathcal{L}_\xi e^\mu_a+\lambda_a{}^b e^\mu_b\,.
\end{eqnarray} 
There are only two invariants, i.e. tensors invariant under $G$ and $J$ transformations, that we can build out of the vielbeins. These are $\tau_\mu$ and $h^{\mu\nu}=\delta^{ab}e^\mu_a e^\nu_b$. This is not enough to construct an affine connection that transforms as \eqref{eq:affinecon}. The reason we cannot build any other invariants is because $v^\mu$ and $h_{\mu\nu}=\delta_{ab}e^a_\mu e^b_\nu$ undergo shift transformations under local Galilean boosts $\lambda^a$ (also known as Milne boosts \cite{Jensen:2014aia}).

\section{The Affine Connection: Part 1 \label{sec:AC1} } 
 
The most general $\Gamma^\rho_{\mu\nu}$ obeying \eqref{eq:conditionGamma1} and \eqref{eq:conditionGamma2} is of the form
\begin{equation}
\Gamma^{\rho}_{\mu\nu} = -v^\rho\partial_\mu\tau_\nu+\frac{1}{2}h^{\rho\sigma}\left(\partial_\mu h_{\nu\sigma}+\partial_\nu h_{\mu\sigma}-\partial_\sigma h_{\mu\nu}\right)+\frac{1}{2}h^{\rho\sigma}Y_{\sigma\mu\nu}\,
\end{equation}
where $h^{\rho\sigma}Y_{\sigma\mu\nu}$ satisfies
\begin{equation}
\left(h^{\lambda\sigma}h^{\rho\nu}+h^{\rho\sigma}h^{\lambda\nu}\right)Y_{\sigma\mu\nu}=0\,.
\end{equation}
It follows that $Y_{\sigma\mu\nu}$ can be written as
\begin{equation}
Y_{\sigma\mu\nu}=\tau_\sigma X^1_{\mu\nu}+\tau_\nu X^2_{\sigma\mu}+X^3_{\sigma\mu\nu}\,,
\end{equation}
where $X^1_{\mu\nu}$ and $X^2_{\sigma\mu}$ and $X^3_{\sigma\mu\nu}=-X^3_{\nu\mu\sigma}$ are arbitrary. We write $X^2_{\sigma\mu}=K_{\sigma\mu}+X^2_{(\sigma\mu)}$ so that $K_{\sigma\mu}=-K_{\mu\sigma}$. Further we write $X^3_{\sigma\mu\nu}=\tau_\mu K_{\sigma\nu}+\tilde X^3_{\sigma\mu\nu}$ so that we can write
\begin{equation}\label{eq:Y}
Y_{\sigma\mu\nu} = \tau_\sigma\left(X^1_{\mu\nu}+X^2_{(\mu\nu)}\right)+\tau_\mu K_{\sigma\nu}+\tau_\nu K_{\sigma\mu}+L_{\sigma\mu\nu}\,,
\end{equation}
where $L_{\sigma\mu\nu}=-L_{\nu\mu\sigma}$ is defined as
\begin{equation}
L_{\sigma\mu\nu}=\tau_\nu X^2_{(\sigma\mu)}-\tau_\sigma X^2_{(\nu\mu)}+\tilde X^3_{\sigma\mu\nu}\,.
\end{equation}
Since $Y_{\sigma\mu\nu}$ is defined as $h^{\rho\sigma}Y_{\sigma\mu\nu}$ we can drop the part in \eqref{eq:Y} that is proportional to $\tau_\sigma$. We thus find the following form for the connection $\Gamma^\rho_{\mu\nu}$
\begin{equation}\label{eq:generalformY}
\Gamma^{\rho}_{\mu\nu} = -v^\rho\partial_\mu\tau_\nu+\frac{1}{2}h^{\rho\sigma}\left(\partial_\mu h_{\nu\sigma}+\partial_\nu h_{\mu\sigma}-\partial_\sigma h_{\mu\nu}\right)+\frac{1}{2}h^{\rho\sigma}\left(\tau_\mu K_{\sigma\nu}+\tau_\nu K_{\sigma\mu}+L_{\sigma\mu\nu}\right)\,.
\end{equation}
The variation of $\Gamma^\rho_{\mu\nu}$ under local Galilean boosts yields
\begin{eqnarray}
\delta_G\Gamma^{\rho}_{\mu\nu} & = & \frac{1}{2}h^{\rho\sigma}\tau_\mu\left(\delta_G K_{\sigma\nu}+\partial_\nu\lambda_\sigma-\partial_\sigma\lambda_\nu\right)+ \frac{1}{2}h^{\rho\sigma}\tau_\nu\left(\delta K_{\sigma\mu}+\partial_\mu\lambda_\sigma-\partial_\sigma\lambda_\mu\right)\\
&&+\frac{1}{2}h^{\rho\sigma}\left(\delta_G L_{\sigma\mu\nu}-\lambda_\sigma\left(\partial_\mu\tau_\nu-\partial_\nu\tau_\mu\right)+\lambda_\mu\left(\partial_\nu\tau_\sigma-\partial_\sigma\tau_\nu\right)+\lambda_\nu\left(\partial_\mu\tau_\sigma-\partial_\sigma\tau_\mu\right)\right)\,,\nonumber
\end{eqnarray} 
where $\lambda_\mu=\lambda_a e^a_\mu$. In section \ref{sec:connection2} we will choose $K_{\mu\nu}$ and $L_{\sigma\mu\nu}$ such that $\delta_G\Gamma^{\rho}_{\mu\nu}=0$.

\section{Local Bargmann Transformations}\label{sec:Bargmann}

It is well known that the Galilean algebra admits a central extension with central element $N$ called the Bargmann algebra. This latter element appears via the commutator $[P_a, G_b]=\delta_{ab}N$. We denote the associated gauge connection by $m_\mu$. Following the same recipe as in section \ref{sec:LocalGalTrafos} with
\begin{eqnarray}
\mathcal{A}_\mu & = & H\tau_{\mu}+P_ae_{\mu}^a+G_a\Omega_{\mu}{}^a+\frac{1}{2}J_{ab}\Omega_{\mu}{}^{ab}+N m_\mu\,,\label{eq:curlABargmann}\\
\Sigma & = & G_a\lambda^a+\frac{1}{2}J_{ab}\lambda^{ab}+N\sigma\,,\\
\mathcal{F}_{\mu\nu} & = & \partial_\mu\mathcal{A}_\nu-\partial_\nu\mathcal{A}_\mu+[{\mathcal{A}}_{\mu}\,,{\mathcal{A}}_{\nu}]\nonumber\\
&=& HR_{\mu\nu}(H)+P_a R_{\mu\nu}{}^a(P)+G_a R_{\mu\nu}{}^a(G)+\frac{1}{2}J_{ab}R_{\mu\nu}{}^{ab}(J)+NR_{\mu\nu}(N)\,,\label{eq:YMcurv}
\end{eqnarray}
we obtain
\begin{equation}\label{eq:trafom}
\bar\delta m_\mu = \mathcal{L}_\xi m_\mu+\partial_\mu\sigma+e_\mu^a\lambda_a\,,
\end{equation}
where $\bar\delta$ is defined in the same way as in \eqref{eq:bardeltaA}. Note that we have an extra parameter $\sigma$ associated with the $N$ transformation. Because $N$ is central, all results of the previous section remain unaffected.

Our primary focus in this section is local Galilean boost invariance. The new field $m_\mu$ is shifted under the $\lambda^a$ transformation and so in combinations such as
\begin{eqnarray}
\hat v^\mu & = & v^\mu-h^{\mu\nu}m_\nu\,,\label{eq:hatv}\\ 
\bar h_{\mu\nu} & = & h_{\mu\nu}-\tau_\mu m_\nu-\tau_\nu m_\mu\,,\label{eq:barh}
\end{eqnarray}
the Galilean boost parameter $\lambda^a$ is cancelled. However we now have two other things to worry about. First of all the new field $m_\mu$ also transforms under a local $U(1)$ transformation with parameter $\sigma$ and secondly we have introduced more than is strictly necessary to have local Galilean invariance. This is because the component 
\begin{equation}\label{eq:tildePhi}
\tilde\Phi=-v^\mu m_\mu+\tfrac{1}{2}h^{\mu\nu}m_\mu m_\nu
\end{equation}
is $G$ invariant (and of course also $J$ invariant). In previous works we have introduced another background field $\chi$, a St\"uckelberg scalar, transforming as $\bar\delta\chi=\mathcal{L}_\xi\chi+\sigma$ so that the combination $M_\mu=m_\mu-\partial_\mu\chi$ is invariant under the local $N$ transformation and replaced everywhere $m_\mu$ by $M_\mu$. Here it will prove convenient, for the sake of comparison with work on HL gravity to postpone this step until later\footnote{In previous work \cite{Hartong:2014oma,Hartong:2014pma,Bergshoeff:2014uea,Hartong:2015wxa} we denoted by $\hat v^\mu$, $\bar h_{\mu\nu}$ and $\tilde\Phi$ the invariants with $m_\mu$ replaced by $M_\mu$. Here we temporarily work with the forms \eqref{eq:hatv}--\eqref{eq:tildePhi} for reasons that will become clear as we go on. We return to our notation from previous works in section \ref{sec:confHLactions}. We also point out that compared to \cite{Hartong:2014oma,Hartong:2014pma,Bergshoeff:2014uea,Hartong:2015wxa} we denote by $m_\mu$ here what was referred to as $\tilde m_\mu$ in these papers and vice versa we denote by $\tilde m_\mu$ here what was denoted by $m_\mu$ in these respective works.\label{fn:notation}}. Hence for now we will work with $m_\mu$ as opposed to $M_\mu$. 

We introduce a new set of Galilean invariant vielbeins: $\tau_\mu$, $\hat e_\mu^a$ whose inverses are $\hat v^\mu$ and $e^\mu_a$ where $\hat e_\mu^a=e_\mu^a-m^a\tau_\mu$ with $m^a=e^{\mu a} m_\mu$. They satisfy the relations 
\begin{equation}
\hat v^\mu\tau_\mu=-1\,,\qquad \hat v^\mu \hat e_\mu^a=0\,,\qquad e^\mu_a\tau_\mu=0\,,\qquad e^\mu_a \hat e_\mu^b=\delta^b_a\,.
\end{equation} 
We also have the completeness relation $e^\mu_a \hat e^a_\nu = \delta^\mu_\nu + \hat v^\mu \tau_\nu$. The introduction of $m^a$ thus leads to the $G$, $J$ invariants $\hat v^\mu$ and
\begin{equation}\label{eq:hath}
\hat h_{\mu\nu}=\delta_{ab}\hat e^a_\mu\hat e^b_\nu=\bar h_{\mu\nu}+2\tau_\mu\tau_\nu\tilde\Phi\,,
\end{equation}
where $\bar h_{\mu\nu}$ is given in \eqref{eq:barh}. The part of $m_\mu$ that is responsible for the Galilean boost invariance is $m^a$ that transforms as (ignoring the $\sigma$ transformation)
\begin{equation}\label{eq:trafoMa}
\bar\delta m^a=\mathcal{L}_\xi m^a+\lambda^a+\lambda^a{}_b m^b\,.
\end{equation}
We can write
\begin{equation}\label{eq:mmu}
m_\mu=e_\mu^a m_a-\frac{1}{2}m_a m^a\tau_\mu+\tilde\Phi\tau_\mu\,,
\end{equation}
where the last term is an invariant.

\section{The Affine Connection: Part 2}\label{sec:connection2}

In section \ref{sec:LocalGalTrafos} we realized the Galilean algebra on the fields $\tau_\mu$, $e_\mu^a$, $\Omega_\mu{}^a$ and $\Omega_\mu{}^{ab}$ or what is the same on $\tau_\mu$, $e_\mu^a$ and $\Gamma^\rho_{\mu\nu}$ where the affine connection obeys \eqref{eq:conditionGamma1} and \eqref{eq:conditionGamma2}. Now that we have introduced a new field $m_\mu$ transforming as in \eqref{eq:trafom} we will see that we can realize the Galilean algebra on a smaller set of fields, namely $\tau_\mu$, $e_\mu^a$ and $m_\mu$. We can also realize the Galilean algebra on $\tau_\mu$, $e_\mu^a$ and $m^a$ with $m^a$ transforming as in \eqref{eq:trafoMa}, i.e. no dependence on $\tilde\Phi$ or realize it on $\tau_\mu$, $e_\mu^a$, $m^a$ and $\tilde\Phi$ which is another way of writing the dependence on $\tau_\mu$, $e_\mu^a$ and $m_\mu$. These different options lead to different choices for the affine connection as we will now discuss. 

The most straightforward way of constructing a $\Gamma^\rho_{\mu\nu}$ that is made out of vielbeins and either i). $m_\mu$ or ii). $m^a$, that obeys \eqref{eq:conditionGamma1} and \eqref{eq:conditionGamma2} and transforms as in \eqref{eq:affinecon}, is to use the invariants $\tau_\mu$, $\bar h_{\mu\nu}$, $\hat v^\mu$,  $h^{\mu\nu}$ and $\tilde\Phi$. The most general connection we can build out of these invariants reads \cite{Hartong:2015wxa}
\begin{equation}\label{eq:GammaTNC}
\Gamma^{\rho}_{\mu\nu} = -\hat v^\rho\partial_\mu\tau_\nu+\frac{1}{2}h^{\rho\sigma}\left(\partial_\mu H_{\nu\sigma}+\partial_\nu H_{\mu\sigma}-\partial_\sigma H_{\mu\nu}\right)\,,
\end{equation}
where $H_{\mu\nu}$ is given by
\begin{equation}\
H_{\mu\nu}=\bar h_{\mu\nu}+\alpha\tau_\mu\tau_\nu\tilde\Phi\,,
\end{equation}
where $\alpha$ is any constant. If we want the connection to depend linearly on $m_\mu$, which is a special case of case i). above, we should take $\alpha=0$. If we wish that the connection is independent of $\tilde\Phi$ as in case ii). we should take $\alpha=2$ because of the identity \eqref{eq:hath} so that $H_{\mu\nu}=\hat h_{\mu\nu}$ where $\hat h_{\mu\nu}$ only depends on $m^a$. For the general case i). i.e. general dependence on $m^a$ and $\tilde\Phi$, we can take any $\alpha$. For case i). with a linear dependence on $m_\mu$ we denote $\Gamma^\rho_{\mu\nu}$ by $\bar\Gamma^\rho_{\mu\nu}$ which is given by
\begin{equation}\label{eq:barGammaTNC}
\bar\Gamma^{\rho}_{\mu\nu} = -\hat v^\rho\partial_\mu\tau_\nu+\frac{1}{2}h^{\rho\sigma}\left(\partial_\mu\bar h_{\nu\sigma}+\partial_\nu \bar h_{\mu\sigma}-\partial_\sigma\bar h_{\mu\nu}\right)\,.
\end{equation}
This form of $\Gamma^\rho_{\mu\nu}$ has been used in \cite{Hartong:2014oma,Hartong:2014pma,Bergshoeff:2014uea,Jensen:2014aia,Bekaert:2014bwa}. The form of $\bar\Gamma^\rho_{\mu\nu}$ corresponds to taking in \eqref{eq:generalformY} the following choices for $K_{\mu\nu}$ and $L_{\sigma\mu\nu}$, namely
\begin{eqnarray}
K_{\mu\nu} & = & \partial_\mu m_\nu-\partial_\nu m_\mu\,,\\ \label{eq:choiceK} 
L_{\sigma\mu\nu} & = & m_\sigma\left(\partial_\mu\tau_\nu-\partial_\nu\tau_\mu\right)-m_\mu\left(\partial_\nu\tau_\sigma-\partial_\sigma\tau_\nu\right)-m_\nu\left(\partial_\mu\tau_\sigma-\partial_\sigma\tau_\mu\right)\,.\label{eq:choiceL}
\end{eqnarray}
For case ii). we denote $\Gamma^\rho_{\mu\nu}$ by $\hat\Gamma^{\rho}_{\mu\nu}$ which reads
\begin{equation}\label{eq:hatGammaTNC}
\hat\Gamma^{\rho}_{\mu\nu} = -\hat v^\rho\partial_\mu\tau_\nu+\frac{1}{2}h^{\rho\sigma}\left(\partial_\mu\hat h_{\nu\sigma}+\partial_\nu \hat h_{\mu\sigma}-\partial_\sigma\hat h_{\mu\nu}\right)\,.
\end{equation}
The two connections $\hat\Gamma^{\rho}_{\mu\nu}$ and $\bar\Gamma^{\rho}_{\mu\nu}$ differ by a tensor as follows from
\begin{equation}\label{eq:relationGammahatbar}
\hat\Gamma^\rho_{\mu\nu}=\bar\Gamma^\rho_{\mu\nu}+\tilde\Phi h^{\rho\sigma}\tau_\nu\left(\partial_\mu\tau_\sigma-\partial_\sigma\tau_\mu\right)+\tilde\Phi h^{\rho\sigma}\tau_\mu\left(\partial_\nu\tau_\sigma-\partial_\sigma\tau_\nu\right)-\tau_\mu\tau_\nu h^{\rho\sigma}\partial_\sigma\tilde\Phi\,.
\end{equation}
In this work it will prove most convenient to use the connection \eqref{eq:hatGammaTNC} as this eases comparison with HL gravity. We stress though that in principle one can take any of the above choices, i.e. any value for $\alpha$, and that the final form of the effective action for HL gravity will take the same form regardless which $\Gamma^\rho_{\mu\nu}$ one chooses as all dependence on $\alpha$ drops out when forming the scalar terms appearing in the action\footnote{This statement can be made more precise in the following way. The Ho\v rava--Lifshitz actions of section \ref{sec:HLactions} such as \eqref{eq:HLaction} take exactly the same form when written in terms of $\bar\Gamma^\rho_{\mu\nu}$ as when expressed in terms of $\hat\Gamma^\rho_{\mu\nu}$. To show this one needs to use the fact that in section \ref{sec:HLactions} it is assumed that $\tau_\mu$ is hypersurface orthogonal which is something that we do not yet impose at this stage. This is because the difference between covariant derivatives using either one or the other connection involves terms proportional to $\tau_\mu$ and since the scalars in the action are formed by using inverse spatial metrics $h^{\mu\nu}$ those terms drop out. The same comments apply when using the general $\alpha$ of \eqref{eq:GammaTNC}, i.e. there is no dependence on $\alpha$.}.

The reader familiar with the literature on NC geometry without torsion might wonder which of these connections relates to the one of NC geometry (as written for example in \cite{Andringa:2010it} and references therein). The usual NC connection is obtained by taking \eqref{eq:barGammaTNC} with $K_{\mu\nu}$ as given in \eqref{eq:choiceK} and $L_{\sigma\mu\nu}=0$ which follows from \eqref{eq:choiceL} and the fact that for NC geometry we have $\partial_\mu\tau_\nu-\partial_\nu\tau_\mu=0$. The possibility of modifying these connections by terms proportional to $\alpha$ was never considered before probably because this breaks manifest local $N$ invariance of the NC connection which depends on $m_\mu$ only via its curl. 

In the presence of torsion the fact that $L_{\sigma\mu\nu}$ is given by \eqref{eq:choiceL} tells us that we have no manifest $N$ invariance of the connection. Further, for no value of $\alpha$ can we find such an invariance. This can be formally solved by adding a new field to the formalism, a St\"uckelberg scalar $\chi$, that cancels the non-invariance. This will be discussed in the next section. One can also take the point of view as in \cite{Jensen:2014aia} that we should just accept the fact that $\bar\Gamma^{\rho}_{\mu\nu}$ is not $N$ invariant as a mere fact and organize couplings to these geometries and fields living on it in such a way that the action is $N$ invariant. This is certainly a viable point of view and agrees with our approach in all these cases where the dependence on $\chi$ can be removed from the theory by field redefinition or simply because it drops out when one tries to make its appearance explicit. 

If one includes $\chi$ there is the benefit that one can also deal with theories that do not have a local $U(1)$ symmetry (because there is an explicit dependence on $\chi$ so that the $U(1)$ invariance disappears in the St\"uckelberg coupling between $m_\mu$ and $\chi$). This is what allows us to use fixed TNC background geometries for both Lifshitz field theories (explicit dependence on $\chi$) as well as Schr\"odinger field theories (no dependence on $\chi$) as discussed in \cite{Hartong:2014pma,Hartong:2015wxa}. The $\chi$ field also allows us, as we will see in section \ref{sec:localU1}, to construct two types of HL actions: those that have a local $U(1)$ symmetry without any dependence on $\chi$ and those that have no local $U(1)$ because $m_\mu$ always appears as $M_\mu=m_\mu-\partial_\mu\chi$.

From now on we will work with \eqref{eq:hatGammaTNC} and simply denote it by $\Gamma^\rho_{\mu\nu}$ unless specifically stated otherwise. With this realization of $\Gamma^\rho_{\mu\nu}$ the other connections $\Omega_\mu{}^{ab}$ and $\Omega_\mu{}^{a}$ are fixed by the vielbein postulates. For an invariant such as $\hat v^\mu$ the covariant derivatives $\nabla_\mu$ and $\mathcal{D}_\mu$ are the same so we can write
\begin{equation}
\nabla_\mu\hat v^\nu=\mathcal{D}_\mu \hat v^\nu=-e^\nu_aD_\mu m^a\,,
\end{equation}
where we used \eqref{eq:invVP1} and \eqref{eq:invVP2} and where $D_\mu m^a$ is given by
\begin{equation}
D_\mu m^a=\partial_\mu m^a-\Omega_\mu{}^a{}_b m^b-\Omega_\mu{}^a\,.
\end{equation}

In this section we focussed on making the affine connection $G$ invariant ($J$ invariance is automatic). It so far is not $N$ invariant. This will be fixed in the next section. We could have made the connection $N$ but not $G$ invariant by taking $K_{\mu\nu}$ as in \eqref{eq:choiceK} and $L_{\sigma\mu\nu}=0$. However in this case we are not achieving anything as the connection without $K_{\mu\nu}$ is also $N$ invariant and so imposing $N$ invariance does not constrain $\Gamma^\rho_{\mu\nu}$. Furthermore since in the transformation of $m_\mu$ the $G$ boost parameter $\lambda^a$ appears without a derivative, whereas the $N$ transformation parameter $\sigma$ appears with a derivative, it is more natural to use $m_\mu$ to make various tensors $G$ invariant.

Using the invariants $\tau_\mu$, $h^{\mu\nu}$, $\hat v^\mu$, $\hat h_{\mu\nu}$ we can build a non-degenerate symmetric rank 2 tensor with Lorentzian signature $g_{\mu\nu}$ that in the case of a relativistic theory we would refer to as a Lorentzian metric. The metric $g_{\mu\nu}$ and its inverse $g^{\mu\nu}$ are given by
\begin{eqnarray}
g_{\mu\nu} & = & -\tau_\mu\tau_\nu+\hat h_{\mu\nu}\,,\label{eq:Lorentzmetric}\\
g^{\mu\nu} & = & -\hat v^\mu\hat v^\nu+h^{\mu\nu}\,,\label{eq:Lorentzinvmetric}
\end{eqnarray}
for which we have
\begin{eqnarray}
g_{\mu\nu}\hat v^\mu & = & \tau_\nu\,,\\
g_{\mu\nu} e^\mu_a & = & \hat e_{\nu a}\,.
\end{eqnarray} 
However the natural Galilean metric structures are $\tau_\mu$ and $h^{\mu\nu}$. For example, as we will see in section \ref{sec:HLactions}, $g_{\mu\nu}$ does not transform homogeneously under local scale transformations and so it is not on the same footing as the Riemannian metric in GR.

\section{Torsion and the St\"uckelberg Scalar}\label{sec:torsion}
 
In the case of gauging the Poincar\'e algebra (appendix \ref{app:gaugingPoincare}) the torsion is the part of $\Gamma^\rho_{\mu\nu}$ that is not fixed by the vielbein postulates. In the case of the Bargmann algebra we see on the other hand that it is the torsion that is fixed, namely it is given by the antisymmetric part of \eqref{eq:GammaTNC}, which reads
 \begin{equation}\label{eq:torsion2}
2\hat\Gamma^\rho_{[\mu\nu]}=-\hat v^\rho\left(\partial_\mu\tau_\nu-\partial_\nu\tau_\mu\right)\,.
\end{equation}
It follows that  the curvature \eqref{eq:curvP} obeys 
\begin{equation}\label{eq:torsion}
R_{\mu\nu}{}^a(P)=m^a\left(\partial_\mu\tau_\nu-\partial_\nu\tau_\mu\right)\,,
\end{equation}
while $R_{\mu\nu}(H)=\partial_\mu\tau_\nu-\partial_\nu\tau_\mu$ is left arbitrary. Using that $R_{\mu\nu}{}^a(P)$ transforms as
\begin{equation}
\bar\delta R_{\mu\nu}{}^a(P)=\mathcal{L}_\xi R_{\mu\nu}{}^a(P)+\lambda^aR_{\mu\nu}(H)+\lambda^a{}_b R_{\mu\nu}{}^b(P)\,,
\end{equation}
we see that the right hand side of \eqref{eq:torsion} transforms in exactly the same way as the left hand side (ignoring the central extension $N$). The right hand side of \eqref{eq:torsion} can be matched to transform correctly under the $N$ transformation by adding the St\"uckelberg scalar $\chi$, i.e. by replacing $m^a$ by $M^a=e^{\mu a}(m_\mu-\partial_\mu\chi)$. This explains why in the presence of torsion, i.e. when $\partial_\mu\tau_\nu-\partial_\nu\tau_\mu\neq 0$, we need the scalar $\chi$. In section \ref{sec:localU1} we will see that there is a similar field in HL gravity whose couplings are precisely obtained by replacing everywhere $m_\mu$ by $M_\mu=m_\mu-\partial_\mu\chi$. From a purely geometrical point of view $\chi$ is needed whenever we have torsion, i.e. when the right hand side of \eqref{eq:torsion} is nonzero to ensure correct transformations under the $N$ generator. 

This does not automatically mean that any field theory coupled to such a background has a nontrivial $\chi$ dependence. There are important cases where the $\chi$ field can be removed by a field redefinition or it simply drops out of the action once one tries to make its appearance explicit. We refer to \cite{Hartong:2015wxa} for field theory examples of the first possibility of removing $\chi$ by field redefinition and to section \ref{sec:localU1} for a HL action that exhibits the second property, namely that $\chi$ drops out.

The $\chi$ field also allows us to make the curvature $R_{\mu\nu}(N)$ appearing in \eqref{eq:YMcurv}, which so far played no role, visible. This goes via the following commutator
\begin{equation}
[D_\mu\,,D_\nu ]\chi=-2\Gamma^\rho_{[\mu\nu]}D_\rho\chi-R_{\mu\nu}(N)\,,
\end{equation}
where $D_{\mu}\chi=\partial_\mu\chi-m_\mu$ and where $R_{\mu\nu}(N)$ is given by
\begin{equation}
R_{\mu\nu}(N)=\partial_\mu m_\nu-\partial_\nu m_\mu-2\Omega_{[\mu}{}^a e_{\nu]a}\,.
\end{equation}
We note that by covariance $D_\mu D_\nu\chi$ involves the Galilean boost connection $\Omega_\mu{}^a$. Using the general form of $\Gamma^\rho_{\mu\nu}$ given in \eqref{eq:generalformY} as well as the vielbein postulate \eqref{eq:VP-2} to express $\Omega_\mu{}^a$ in terms of $\Gamma^\rho_{\mu\nu}$ we obtain
\begin{equation}
R_{\mu\nu}(N)=\partial_\mu m_\nu-\partial_\nu m_\mu-K_{\mu\nu}+v^\sigma L_{\sigma[\mu\nu]}\,.
\end{equation}
For the choice $\Gamma^\rho_{\mu\nu}=\bar \Gamma^\rho_{\mu\nu}$ \eqref{eq:barGammaTNC}, i.e. for $K_{\mu\nu}$ and $L_{\sigma\mu\nu}$ as in \eqref{eq:choiceK} and \eqref{eq:choiceL} we find
\begin{equation}
R_{\mu\nu}(N)=v^\sigma m_\sigma\left(\partial_\mu\tau_\nu-\partial_\nu\tau_\mu\right)\,.
\end{equation}
This curvature constraint is in agreement with the curvature constraint \eqref{eq:torsion} because it obeys the transformation rule for the curvatures under Galilean boosts which according to \eqref{eq:trafo2} and \eqref{eq:trafo3} reads $\delta_G R_{\mu\nu}(N)=\lambda^a R_{\mu\nu a}(P)$. Again in order that $R_{\mu\nu}(N)$ remains inert under $N$ transformations in the presence of torsion we need to replace in $\bar\Gamma^\rho_{\mu\nu}$ (more precisely in $L_{\sigma\mu\nu}$ as given in \eqref{eq:choiceL}) $m_\mu$ by $M_\mu=m_\mu-\partial_\mu\chi$. The field $\chi$ is an essential part of NC geometry with torsion. 

The curvature constraints derived here by using the approach of section \ref{sec:LocalGalTrafos} agree with \cite{Andringa:2010it} where the torsionless case was studied. The analysis of sections \ref{sec:LocalGalTrafos}--\ref{sec:torsion} can thus be viewed as adding torsion to the gauging of the Bargmann algebra (without adding dilatations as in \cite{Bergshoeff:2014uea}). By employing the relation \eqref{eq:relationGammahatbar} between $\bar\Gamma^\rho_{\mu\nu}$ and $\hat\Gamma^\rho_{\mu\nu}$ we can find the curvature constraint for $R_{\mu\nu}(N)$ that relates to this choice of affine connection. The curvature constraint \eqref{eq:torsion} is the same for all affine connections \eqref{eq:GammaTNC}.

Following \cite{Christensen:2013lma,Christensen:2013rfa} we distinguish three cases for the torsion \eqref{eq:torsion2}:
\begin{enumerate}
\item No torsion: $\partial_\mu\tau_\nu-\partial_\nu\tau_\mu=0$ which is called Newton--Cartan (NC) geometry.
\item Twistless torsion: $\tau_{[\mu}\partial_\nu\tau_{\rho]}=0$ which means that $\tau_\mu$ is hypersurface orthogonal and is called twistless torsional Newton--Cartan (TTNC) geometry because it is equivalent to \eqref{eq:TTNC} which states that the twist tensor is zero.
\item No constraint on $\tau_\mu$ which is a novel extension of Newton--Cartan (TNC) geometry.
\end{enumerate} 
TTNC geometry goes back to \cite{Julia:1994bs} but in that work a conformal rescaling was done to go to a frame in which there is no torsion. The benefit of adding torsion to the formalism was first considered in \cite{Christensen:2013lma,Christensen:2013rfa} including the case with no constraint on $\tau_\mu$.

We will see below that making NC and TTNC geometries dynamical corresponds to projectable and non-projectable HL gravity. In this work we will always assume that we are dealing with TTNC geometry which contains NC geometry as a special case. 

For twistless torsional Newton--Cartan (TTNC) geometry we have by definition
\begin{equation}\label{eq:TTNC}
h^{\mu\rho}h^{\nu\sigma}\left(\partial_\rho\tau_\sigma-\partial_\sigma\tau_\rho\right)=0\,.
\end{equation}
This implies that the geometry induced on the slices to which $\tau_\mu$ is hypersurface orthogonal is described by (torsion free) Riemannian geometry.

To make contact with the HL literature concerning non-projectable HL gravity it will prove convenient to define a vector $a_\mu$ as follows
\begin{equation}\label{eq:defa}
a_\mu =  \mathcal{L}_{\hat v}\tau_\mu\,.
\end{equation}
In section \ref{sec:parametrization} we will exhibit a coordinate parameterization of $a_\mu$ (see equations \eqref{eq:at} and \eqref{eq:ai}) that will appear more familiar in the context of HL gravity, where this becomes the acceleration of the unit vector field orthogonal to equal time slices.

For TTNC we have the following useful identities
\begin{eqnarray}
h^{\mu\rho}h^{\nu\sigma}\left(\partial_\rho a_\sigma-\partial_\sigma a_\rho\right) & = & h^{\mu\rho}h^{\nu\sigma}\left(\nabla_\rho a_\sigma-\nabla_\sigma a_\rho\right)=0\,,\label{eq:TTNCidentity1}\\
\partial_\mu \tau_\nu-\partial_\nu \tau_\mu & = & a_\mu\tau_\nu-a_\nu\tau_\mu \,.\label{eq:TTNCidentity2}
\end{eqnarray}
The first of these two identities tells us that the twist tensor (the left hand side) vanishes which is why we refer to the geometry as twistless torsional NC geometry. The last identity tells us that $a_\mu$ describes the TTNC torsion. We will thus refer to it as the torsion vector.

\section{Curvatures}\label{sec:curvatures}
 
We start by giving some basic properties of the Riemann tensor \eqref{eq:Riemann} with connection \eqref{eq:hatGammaTNC}. Using that
\begin{equation}
\Gamma^\rho_{\mu\rho}=e^{-1}\partial_\mu e\,,
\end{equation}
where $e=\text{det}\left(\tau_\mu\,,e_\mu^a\right)$, we obtain
\begin{equation}
R_{\mu\nu\rho}{}^\rho=0\,.
\end{equation}
Note that because of torsion we have
\begin{equation}\label{eq:traceGamma}
\Gamma^\rho_{\rho\mu}=e^{-1}\partial_\mu e-\hat v^\rho\left(\partial_\rho\tau_\mu-\partial_\mu\tau_\rho\right)\,.
\end{equation}
From the definition of the Riemann tensor and our choice of connection we can derive the identity
\begin{eqnarray}
3R_{[\mu\nu\sigma]}{}^\rho & = &\left(\nabla_\mu\hat v^\rho\right)\left(\partial_\nu\tau_\sigma-\partial_\sigma\tau_\nu\right)+\left(\nabla_\sigma\hat v^\rho\right)\left(\partial_\mu\tau_\nu-\partial_\nu\tau_\mu\right)\nonumber\\
&&+\left(\nabla_\nu\hat v^\rho\right)\left(\partial_\sigma\tau_\mu-\partial_\mu\tau_\sigma\right)\,.\label{eq:totallyASpartRiem}
\end{eqnarray} 
The trace of this equation gives us the antisymmetric part of the Ricci tensor $R_{\mu\nu}=R_{\mu\rho\nu}{}^\rho$.

The covariant derivative of $\hat v^\mu$ is essentially the extrinsic curvature. Using the connection \eqref{eq:hatGammaTNC} we find the identity 
\begin{equation}\label{eq:covderhatv}
\nabla_\mu\hat v^\rho=-e^\rho_aD_\mu m^a=-h^{\rho\sigma}K_{\mu\sigma}\,,
\end{equation}
where the extrinsic curvature is defined as
\begin{equation}\label{eq:extrinsiccurv}
K_{\mu\nu}=-\frac{1}{2}\mathcal{L}_{\hat v}\hat h_{\mu\nu}\,.
\end{equation}

For TTNC geometries the antisymmetric part of the Ricci tensor is given by
\begin{equation}
2R_{\rho[\mu\nu]}{}^\rho=\left(\nabla_\rho\hat v^\rho\right)\left(a_\mu\tau_\nu-a_\nu\tau_\mu\right)+\hat v^\rho\left(\tau_\nu\nabla_\mu a_\rho-\tau_\mu\nabla_\nu a_\rho\right)\,,
\end{equation}
using \eqref{eq:TTNCidentity2} and \eqref{eq:totallyASpartRiem}. We can also derive a TTNC Bianchi identity that reads
\begin{equation}
3\nabla_{[\lambda}R_{\mu\nu]\sigma}{}^\kappa=2\Gamma^\rho_{[\mu\nu]}R_{\lambda\rho\sigma}{}^\kappa+2\Gamma^\rho_{[\lambda\mu]}R_{\nu\rho\sigma}{}^\kappa+2\Gamma^\rho_{[\nu\lambda]}R_{\mu\rho\sigma}{}^\kappa\,.
\end{equation}
Contracting $\lambda$ and $\kappa$ and the remaining indices with $\hat v^\mu h^{\nu\sigma}$ leads to the identity
\begin{eqnarray}
0 & = & e^{-1}\partial_\mu\left(e\hat v^\nu h^{\mu\sigma}R_{\nu\kappa\sigma}{}^\kappa\right)-\frac{1}{2}e^{-1}\partial_\mu\left(e\hat v^\mu h^{\nu\sigma}R_{\nu\kappa\sigma}{}^\kappa\right)+h^{\mu\rho}h^{\nu\sigma}K_{\rho\sigma}R_{\mu\kappa\nu}{}^\kappa\nonumber\\
&&-\frac{1}{2}h^{\mu\nu}K_{\mu\nu}h^{\rho\sigma}R_{\rho\kappa\sigma}{}^\kappa\,,
\end{eqnarray}
where we used \eqref{eq:traceGamma} and \eqref{eq:covderhatv}. Since we will mostly work in 2+1 dimensions we focus on what happens in that case. Using \eqref{eq:decompRiemann} we find
\begin{equation}\label{eq:3DBI}
e^{-1}\partial_\mu\left(e\hat v^\nu h^{\mu\sigma}R_{\nu\kappa\sigma}{}^\kappa\right)+\frac{1}{2}e^{-1}\partial_\mu\left(e\hat v^\mu \mathcal{R}\right)=0\,,
\end{equation}
where we used that in 2 spatial dimensions 
\begin{equation}\label{eq:2DRiemann}
R_{abcd}(J)=\frac{1}{2}\left(\delta_{ac}\delta_{bd}-\delta_{ad}\delta_{bc}\right)\mathcal{R}\,.
\end{equation}

\section{Coordinate (ADM) Parametrizations}\label{sec:parametrization}

Even though we treat the NC fields $\tau_\mu$ and $\hat h_{\mu\nu}$ as independent we can parametrize them in such a way that $g_{\mu\nu}$ in \eqref{eq:Lorentzmetric} is written in an ADM decomposition. Writing
\begin{equation}
ds^2=g_{\mu\nu}dx^\mu dx^\nu=-N^2 dt^2+\gamma_{ij}\left(dx^i+N^idt\right)\left(dx^j+N^jdt\right)\,,
\end{equation}
leads to 
\begin{eqnarray}
\hat h_{tt} & = & \gamma_{ij}N^i N^j+\tau_t^2-N^2\,,\\
\hat h_{ti} & = & \gamma_{ij}N^j+\tau_i\tau_t\,,\\
\hat h_{ij} & = & \gamma_{ij}+\tau_i\tau_j\,.
\end{eqnarray}
For the inverse metric \eqref{eq:Lorentzinvmetric} the ADM decomposition reads
\begin{eqnarray}
g^{tt} & = & -N^{-2}\,,\\
g^{ti} & = & N^iN^{-2}\,,\\
g^{ij} & = & \gamma^{ij}-N^iN^jN^{-2}\,.
\end{eqnarray}
From this we conclude that
\begin{eqnarray}
h^{tt} & = & -N^{-2}+\hat v^t\hat v^t\,,\\
h^{ti} & = & N^iN^{-2}+\hat v^t\hat v^i\,,\\
h^{ij} & = & \gamma^{ij}-N^iN^jN^{-2}+\hat v^i\hat v^j\,.
\end{eqnarray}
The choice \eqref{eq:TTNC} implies that $\tau_\mu$ is hypersurface orthogonal, i.e.
\begin{equation}\label{eq:TTNCtau}
\tau_{\mu}=\psi\partial_\mu\tau\,.
\end{equation}
If we fix our choice of coordinates such that $\tau=t$ we obtain 
\begin{equation}\label{eq:foliation}
\tau_i=0\,.
\end{equation}

Using that $\tau_\mu h^{\mu\nu}=0$ and \eqref{eq:foliation} we obtain $h^{tt}=h^{ti}=0$ as well as $\hat h_{ti} = \gamma_{ij}N^j$ and $\hat h_{ij} = \gamma_{ij}$. Further using that $h^{\mu\rho}\hat h_{\nu\rho}=\delta^\mu_\nu+\hat v^\mu\tau_\nu$ we find $h^{ij} = \gamma^{ij}$. This in turn tells us that $\hat v^i = N^iN^{-1}$, so that $h^{tt}=h^{ti}=0$ leads to $\hat v^t = -N^{-1}$. Since $\hat v^\mu\tau_\mu=-1$ we also obtain $\tau_t=\psi=N$ so that $\hat h_{tt} = \gamma_{ij}N^i N^j$. Since $h^{tt}=h^{ti}=0$ we also have $\hat v^t=v^t=-N^{-1}$ which in turn tells us that $h_{ti}=h_{tt}=0$, so that we find 
\begin{equation}\label{eq:mi}
m_i=-\gamma_{ij}\frac{N^j}{N}\,.
\end{equation}
Furthermore we have $h_{ij}=\gamma_{ij}$ and $v^i=0$. For the time component of $m_\mu$ we obtain
\begin{equation}\label{eq:mt}
m_t=-\frac{1}{2N}\gamma_{ij}N^i N^j+N\tilde\Phi\,,
\end{equation}
where we used \eqref{eq:mmu} or alternatively \eqref{eq:hath} and \eqref{eq:barh}. In general $\tau_t=N=N(t,x)$ so that we are dealing with non-projectable HL gravity. Projectable HL gravity corresponds to $N=N(t)$ which is precisely what we get when we impose $\partial_\mu\tau_\nu-\partial_\nu\tau_\mu=0$.

In these coordinates the torsion vector \eqref{eq:defa} reduces to
\begin{eqnarray}
a_t & = & N^i a_i\,,\label{eq:at}\\
a_i & = & N^{-1}\partial_i N\,,\label{eq:ai}
\end{eqnarray}
which contains no time derivatives. The determinant $e$ in this parametrization is given by $N\sqrt{\gamma}$ where $\gamma$ is the determinant of $\gamma_{ij}$ so that using \eqref{eq:traceGamma} we find $\Gamma^\rho_{\rho i}=\partial_i\log\sqrt{\gamma}$ making an object such as $\nabla_\mu (h^{\mu\nu}X_\nu)$ a $\gamma$-covariant spatial divergence.

The number of components in $g_{\mu\nu}$ in $d+1$ space-time dimensions is $(d+1)(d+2)/2$ whereas the total number of components in $\tau_\mu$ and $\hat h_{\mu\nu}$ is $(d+1)(d+2)/2+d+1-1$ where the extra $d+1$ originate from $\tau_\mu$ and the $-1$ comes from the fact that $\hat h_{\mu\nu}=\delta_{ab}\hat e^a_\mu\hat e^b_\nu$ so that it has zero determinant. If we furthermore use the fact that $\tau_\mu$ is hypersurface orthogonal, i.e. $\tau_\mu=\psi\partial_\mu\tau$, we can remove another $d-1$ components ending up with  $(d+1)(d+2)/2+1$ which is one component more than we have in $g_{\mu\nu}$. If we next restrict to coordinate systems for which $\tau=t$ we obtain the same number of components in the ADM decomposition as we have for our TTNC geometry without $\tilde\Phi$. Later we will see what the scalars $\tilde\Phi$ and the St\"uckelberg scalar $\chi$ (mentioned below \eqref{eq:tildePhi}) correspond to in the context of HL gravity. 
 
This counting exercise also shows that in general for arbitrary $\tau_\mu$ TNC gravity is much more general than HL gravity. We leave the study of this more general case for future research. Here we restrict to a hypersurface orthogonal $\tau_\mu$.
 
We thus see that the field $\tau_\mu$ describes many properties that we are familiar with from the HL literature. For example the TTNC form of $\tau_\mu$ in \eqref{eq:TTNCtau} agrees with the Khronon field of \cite{Blas:2010hb}. More precisely the Khronon field $\varphi$ of \cite{Blas:2010hb} corresponds to what we call $\tau$ and what is called $u_\mu$ in \cite{Blas:2010hb} corresponds to what we call $\tau_\mu$. Further the torsion field $a_i$ that we defined via \eqref{eq:defa} and that has the parametrization \eqref{eq:ai} agrees with the same field appearing in \cite{Blas:2010hb} where it is referred to as the acceleration vector. We will now show that the generic action describing dynamical TTNC geometries agrees on the nose with the most general HL actions appearing in the literature.

\section{Ho\v rava--Lifshitz Actions}\label{sec:HLactions}

We will consider the dynamics of geometries described by $\tau_\mu$, $e_\mu^a$ and $m^a$ (in the next section we will add $\tilde\Phi$ and $\chi$) by ensuring manifest $G$ and $J$ invariance and by constructing in a systematic manner (essentially a derivative expansion) an action for these fields. Since we demand manifest $G$ and $J$ invariance the generic theory will be described by the independent fields $\tau_\mu$ and $\hat h_{\mu\nu}$ and derivatives thereof.

For simplicity we will work with twistless torsion and in $2$ spatial dimensions with $1<z\le 2$. It is straightforward to consider higher dimensions. We will do this in section \ref{sec:confHLactions} where we treat the conformal case. A convenient way to organize the terms in the action is according to their dilatation weight. The dilatation weights of the invariants are given in table \ref{table:dimensionsinvariants} where $e$ is the determinant of the matrix $(\tau_\mu\,, e_\mu^a)$.
\begin{table}[h!]
      \centering
      \begin{tabular}{|c|c|c|c|c|c||c|c|}
      \hline
$G$, $J$ invariant & $\tau_\mu$ & $\hat h_{\mu\nu}$ & $\hat v^\mu$ & $h^{\mu\nu}$ & $e$ & $\tilde\Phi$ & $\chi$ \\
  \hline
 dilatation weight & $-z$ & $-2$ & $z$ & $2$ & $-(z+2)$ & $2(z-1)$ & $z-2$  \\
         \hline
           \end{tabular}
      \caption{Dilatation weights of the $G,J$ invariants.}\label{table:dimensionsinvariants}
\end{table} 
The assignment of these dilatation weights to the TNC fields is consistent with the fact that adding dilatations to the Bargmann algebra leads to the Schr\"odinger algebra for general $z$ \cite{Hartong:2014oma,Bergshoeff:2014uea}. These assignments agree with \cite{Horava:2009uw}. If we choose the foliation as in the previous section with $\tau_i=0$ and assign the length dimensions $z$ and $1$ to the coordinates $t$ and $x^i$, respectively, we obtain that $[\tau_t]=[N]=L^0$, $[N^i]=L^{1-z}$ and $\gamma_{ij}=L^0$. Note that in table \ref{table:dimensionsinvariants} we do not assign any dilatation weights to the coordinates. In the last two columns we have added the scalars $\tilde\Phi$ and $\chi$ that will not be used in this section but that will appear in the following sections. Even though the fields transform in representations of the Schr\"odinger algebra this does not mean that this a local symmetry of the action. This case will be studied in section \ref{sec:confHLactions} leading to conformal HL actions.

There are three ways of building derivative terms, namely by i). employing the torsion tensor \eqref{eq:torsion2}, ii). taking covariant derivatives of $\tau_\mu$ and $\hat h_{\mu\nu}$ as well as covariant derivatives of the torsion tensor and iii). by building scalars out of the $G$, $J$ (and later $N$) invariants and the curvature tensor $R_{\mu\nu\sigma}{}^\rho$. Option one amounts to using the combination $\partial_\mu\tau_\nu-\partial_\nu\tau_\mu$ which because of our choice \eqref{eq:TTNC} means that the only relevant component is the one obtained by contracting $\partial_\mu\tau_\nu-\partial_\nu\tau_\mu$ with $\hat v^\mu$ which equals the Lie derivative of $\tau_\nu$ along $\hat v^\mu$. In other words we can employ the vector $a_\mu$ defined in \eqref{eq:defa}. Option two reduces to just the covariant derivative of $\hat h_{\mu\nu}$ and $a_\mu$ because of what was just said about the torsion tensor and the fact that $\nabla_\mu\tau_\nu=0$. If we contract $\nabla_\rho\hat h_{\mu\nu}$ with $h^{\lambda\mu}h^{\kappa\nu}$ we obtain zero because of the fact that $\nabla_\rho h^{\lambda\kappa}=0$. This means that the only relevant part of 
$\nabla_\rho\hat h_{\mu\nu}$ is obtained by contracting it with one $\hat v^\mu$ (two would give zero). Since we have $\hat v^\mu\nabla_\rho\hat h_{\mu\nu}=-\hat h_{\mu\nu}\nabla_\rho\hat v^\mu$ we can reduce option 2 to taking covariant derivatives of $\hat v^\mu$ and $h^{\mu\nu}a_\nu$ (note that $\hat v^\mu a_\mu=0$). Because of the identity \eqref{eq:covderhatv} or what is the same
\begin{equation}
\hat h_{\nu\rho}\nabla_\mu\hat v^\rho=-K_{\mu\nu}\,,
\end{equation}
the extrinsic curvature can be viewed as the covariant derivative of $\hat v^\mu$. Options 1 and 2 thus amount to taking the vectors $h^{\mu\nu}a_\nu$ and $\hat v^\mu$ as well as products thereof and to form scalar invariants by acting on these tensors with covariant derivatives and/or (products of) $a_\mu$. We will now first classify these terms before discussing option 3.

We will classify all terms that are at most second order in time derivatives and that have no dilatation weights 
higher than $z+2$ (which is the negative of the dilatation weight of $e$). In other words we only consider relevant and marginal couplings. The only terms containing time derivatives are extrinsic curvature terms which as we observed are covariant derivatives of $\hat v^\mu$. In the previous section we observed that $a_\mu$ does not contain any time derivatives, see equations \eqref{eq:at} and \eqref{eq:ai}. We start by writing down all products of $\hat v^\mu$ and $h^{\mu\nu}a_\nu$ that have dilatation weight at most $z+2$, taking into consideration that we restrict our attention to the range $1<z\le 2$. The possibilities are
\begin{equation}\label{eq:allowedterms}
\begin{array}{rcl}
\hat v^\mu &\phantom{a}\quad& z\\
h^{\mu\nu}a_\nu &\quad& 2\\
\hat v^\mu\hat v^\nu &\quad& 2z\\
\hat v^\mu h^{\nu\rho}a_\rho &\quad& z+2\\
h^{\mu\rho}a_\rho h^{\nu\sigma}a_\sigma &\quad& 4\,,
\end{array}
\end{equation}
where the dilatation weights are indicated in the second column. Terms with weight 4 are only relevant for the case $z=2$. We now hit these terms with $\nabla_\mu$ and $a_\mu$ in all possible ways to form scalars. This does not change the dilatation weights because both $\nabla_\mu$ and $a_\mu$ have weight zero. Keeping in mind that $\hat v^\mu a_\mu=0$ the first two terms in \eqref{eq:allowedterms} give rise to the following scalars
\begin{equation}\label{eq:scalars1}
\begin{array}{rcl}
\nabla_\mu\hat v^\mu &\phantom{a}\quad& z\\
\nabla_\mu\left(h^{\mu\nu}a_\nu\right) &\quad& 2\\
h^{\mu\nu}a_\mu a_\nu &\quad& 2\,.
\end{array}
\end{equation}
Using \eqref{eq:traceGamma} we have the identity
\begin{equation}
\nabla_\mu X^\mu=e^{-1}\partial_\mu\left(eX^\mu\right)-a_\mu X^\mu\,.
\end{equation}
It follows that the first term in \eqref{eq:scalars1} is a total derivative and the second equals minus the third up to a total derivative. Nevertheless these quantities will be useful as they can be multiplied with a Ricci-type curvature scalar as we will see later. We now focus on the last three terms in \eqref{eq:allowedterms}. There are two free indices so we can contract them with $a_\mu a_\nu$, $a_\mu\nabla_\nu$ and $\nabla_\mu\nabla_\nu$. Using two $a_\mu$'s only leads to one possibility which is
\begin{equation}
\left(h^{\mu\nu}a_\mu a_\nu\right)^2\qquad 4\,.
\end{equation}
Contracting the term $\hat v^\mu\hat v^\nu$ with $a_\mu\nabla_\nu$ gives always zero because we have $a_\mu\hat v^\mu\nabla_\nu\hat v^\nu=0$ and $a_\mu\left(\nabla_\nu\hat v^\mu\right)\hat v^\nu=0$ where the last identity follows from \eqref{eq:covderhatv}. Doing the same with the term $\hat v^\mu h^{\nu\rho}a_\rho$ in the list \eqref{eq:allowedterms} we obtain the following three allowed scalars
\begin{equation}\label{eq:scalars2}
\begin{array}{rcl}
h^{\nu\rho}a_\nu a_\rho\nabla_\mu\hat v^\mu &\phantom{a}\quad& z+2\\
h^{\nu\rho}a_\rho a_\mu\nabla_\nu\hat v^\mu &\phantom{a}\quad& z+2\\
a_\nu \hat v^\mu\nabla_\mu\left(h^{\nu\rho}a_\rho\right) &\phantom{a}\quad& z+2\,.
\end{array}
\end{equation}
However, because of the identity
\begin{equation}
a_\nu \hat v^\mu\nabla_\mu\left(h^{\nu\rho}a_\rho\right)=\frac{1}{2}\hat v^\mu\nabla_\mu\left(h^{\nu\rho}a_\nu a_\rho\right)=-\frac{1}{2}h^{\nu\rho}a_\nu a_\rho\nabla_\mu\hat v^\mu+\text{tot.der.}\,,
\end{equation}
the last of these three terms brings nothing new. Finally the last term in the list \eqref{eq:allowedterms} when contracted with one $a_\mu$ and one $\nabla_\nu$ provides two more scalars, namely
\begin{equation}\label{eq:scalars3}
\begin{array}{rcl}
h^{\mu\rho}a_\mu a_\rho\nabla_\nu\left(h^{\nu\sigma}a_\sigma\right) &\phantom{a}\quad& 4\\
h^{\mu\rho}a_\rho a_\nu\nabla_\mu\left(h^{\nu\sigma}a_\sigma\right) &\phantom{a}\quad& 4\,.
\end{array}
\end{equation}
The second term however brings nothing new because of the identity
\begin{equation}
h^{\mu\rho}a_\rho a_\nu\nabla_\mu\left(h^{\nu\sigma}a_\sigma\right)=-\frac{1}{2}\left(h^{\mu\nu}a_\mu a_\nu\right)^2-\frac{1}{2}h^{\mu\rho}a_\mu a_\rho\nabla_\nu\left(h^{\nu\sigma}a_\sigma\right)+\text{tot.der.}\,.
\end{equation}
Finally we can contract the last three terms in \eqref{eq:allowedterms} with two $\nabla_\mu$'s leading to the following set of scalars
\begin{equation}\label{eq:scalars4}
\begin{array}{rcl}
\nabla_\mu\hat v^\mu\nabla_\nu\hat v^\nu &\phantom{a}\quad& 2z\\
\nabla_\nu\hat v^\mu\nabla_\mu\hat v^\nu &\quad& 2z\\
\nabla_\mu\hat v^\mu \nabla_\nu\left(h^{\nu\rho}a_\rho\right) &\quad& z+2\\
\nabla_\nu\hat v^\mu \nabla_\mu\left(h^{\nu\rho}a_\rho\right) &\quad& z+2\\
\nabla_\mu\left(h^{\mu\rho}a_\rho\right)\nabla_\nu\left(h^{\nu\sigma}a_\sigma\right) &\quad& 4\\
\nabla_\nu\left(h^{\mu\rho}a_\rho\right)\nabla_\mu\left(h^{\nu\sigma}a_\sigma\right) &\quad& 4\,.
\end{array}
\end{equation}
There is one other set of scalar terms containing two covariant derivatives that follow by acting with $a_\mu\square$ where $\square=h^{\rho\sigma}\nabla_\rho\nabla_\sigma$, which is a dimension 2 operator, on the first two terms appearing in the list \eqref{eq:allowedterms}. This leads to 
\begin{equation}\label{eq:scalars4prime}
\begin{array}{rcl}
a_\mu\square \hat v^\mu & \phantom{a}\quad& 2+z\\
a_\mu\square \left(h^{\mu\nu}a_\nu\right) & \quad& 4\,.
\end{array}
\end{equation}
Both of these however give nothing new as can be shown by partial integration and upon using the TTNC identity \eqref{eq:TTNCidentity1}.

We are left with the possibility to add scalar curvature terms. To this end we first introduce a Ricci-type scalar curvature $\mathcal{R}$ defined as
\begin{equation}
\mathcal{R}=-h^{\mu\nu}R_{\mu\rho\nu}{}^\rho\,,
\end{equation}
which has dilatation weight 2. Using the scalars \eqref{eq:scalars1} we can thus build the following list of scalar terms
\begin{equation}\label{eq:scalars5}
\begin{array}{rcl}
\mathcal{R}&\phantom{a}\quad& 2\\
\mathcal{R}\nabla_\mu\hat v^\mu &\phantom{a}\quad& z+2\\
\mathcal{R}^2&\phantom{a}\quad& 4\\
\mathcal{R}\nabla_\mu\left(h^{\mu\nu}a_\nu\right) &\quad& 4\\
\mathcal{R}h^{\mu\nu}a_\mu a_\nu &\quad& 4\,.
\end{array}
\end{equation}
The last term in \eqref{eq:scalars5} makes it possible to remove $\nabla_\mu\left(h^{\mu\rho}a_\rho\right)\nabla_\nu\left(h^{\nu\sigma}a_\sigma\right)$ from the list \eqref{eq:scalars4}. This is due to the identity
\begin{eqnarray}
\nabla_\mu\left(h^{\mu\rho}a_\rho\right)\nabla_\nu\left(h^{\nu\sigma}a_\sigma\right) & = & \nabla_\nu\left(h^{\mu\rho}a_\rho\right)\nabla_\mu\left(h^{\nu\sigma}a_\sigma\right)-\frac{1}{2}\left(h^{\mu\nu}a_\mu a_\nu\right)^2\nonumber\\
&&-\frac{3}{2}h^{\mu\rho}a_\mu a_\rho\nabla_\nu\left(h^{\nu\sigma}a_\sigma\right)-\frac{1}{2}\mathcal{R}h^{\rho\sigma}a_\rho a_\sigma+\text{tot.derv.}\,,
\end{eqnarray}
where we used \eqref{eq:nablacommutator}, \eqref{eq:decompRiemann}, \eqref{eq:traceGamma}, \eqref{eq:2DRiemann} and partial integrations.

In $d=2$ spatial dimensions there are no other curvature invariants other than $\mathcal{R}$. The reason is that all curvature invariants built out of the tensor $R_{\mu\nu\sigma}{}^\rho$ only involve the spatial Riemann tensor $R_{\mu\nu}{}^{ab}(J)$. The tensor $R_{abcd}=e^\mu_a e^\nu_b R_{\mu\nu}{}_{cd}(J)$ has the same symmetry properties as the Riemann tensor of a $d$-dimensional Riemannian geometry. Hence since here $d=2$ the only component is the Ricci scalar $\mathcal{R}$. Any other term involving the curvature tensor contracted with $\hat v^\mu$ or $h^{\mu\nu}a_\nu$ can be written as a combination of terms we already classified using \eqref{eq:nablacommutator} and other identities.

We thus conclude that for $d=2$ and $1<z\le 2$ the scalar terms that can appear in the action are 
\begin{equation}\label{eq:scalars6}
\begin{array}{rcl}
h^{\mu\nu}a_\mu a_\nu &\phantom{a}\quad& 2\\
\mathcal{R}&\phantom{a}\quad& 2\\
\nabla_\mu\hat v^\mu\nabla_\nu\hat v^\nu &\phantom{a}\quad& 2z\\
\nabla_\nu\hat v^\mu\nabla_\mu\hat v^\nu &\quad& 2z\\
h^{\nu\rho}a_\nu a_\rho\nabla_\mu\hat v^\mu &\phantom{a}\quad& z+2\\
h^{\nu\rho}a_\rho a_\mu\nabla_\nu\hat v^\mu &\phantom{a}\quad& z+2\\
\nabla_\mu\hat v^\mu \nabla_\nu\left(h^{\nu\rho}a_\rho\right) &\quad& z+2\\
\nabla_\nu\hat v^\mu \nabla_\mu\left(h^{\nu\rho}a_\rho\right) &\quad& z+2\\
\mathcal{R}\nabla_\mu\hat v^\mu &\phantom{a}\quad& z+2\\
\left(h^{\mu\nu}a_\mu a_\nu\right)^2&\phantom{a}\quad& 4\\
h^{\mu\rho}a_\mu a_\rho\nabla_\nu\left(h^{\nu\sigma}a_\sigma\right) &\phantom{a}\quad& 4\\
\nabla_\nu\left(h^{\mu\rho}a_\rho\right)\nabla_\mu\left(h^{\nu\sigma}a_\sigma\right) &\quad& 4\\
\mathcal{R}^2&\phantom{a}\quad& 4\\
\mathcal{R}\nabla_\mu\left(h^{\mu\nu}a_\nu\right) &\quad& 4\\
\mathcal{R}h^{\mu\nu}a_\mu a_\nu &\quad& 4\,.
\end{array}
\end{equation}
Consequently, we arrive at the action
\begin{eqnarray}
S & = & \int d^3x e\left[c_1 h^{\mu\nu}a_\mu a_\nu+c_2 \mathcal{R}+c_3 \nabla_\mu\hat v^\mu\nabla_\nu\hat v^\nu+c_4 \nabla_\nu\hat v^\mu\nabla_\mu\hat v^\nu+c_5 h^{\nu\rho}a_\nu a_\rho\nabla_\mu\hat v^\mu\right.\nonumber\\
&&\left.+c_6 h^{\nu\rho}a_\rho a_\mu\nabla_\nu\hat v^\mu+c_7 \nabla_\mu\hat v^\mu \nabla_\nu\left(h^{\nu\rho}a_\rho\right) +c_8\nabla_\nu\hat v^\mu \nabla_\mu\left(h^{\nu\rho}a_\rho\right)+c_9\mathcal{R}\nabla_\mu\hat v^\mu\right.\nonumber\\
&&\left.+\delta_{z,2}\left[c_{10}\left(h^{\mu\nu}a_\mu a_\nu\right)^2+c_{11}h^{\mu\rho}a_\mu a_\rho\nabla_\nu\left(h^{\nu\sigma}a_\sigma\right)+c_{12}\nabla_\nu\left(h^{\mu\rho}a_\rho\right)\nabla_\mu\left(h^{\nu\sigma}a_\sigma\right)\right.\right.\nonumber\\
&&\left.\left.+c_{13}\mathcal{R}^2+c_{14}\mathcal{R}\nabla_\mu\left(h^{\mu\nu}a_\nu\right)+c_{15}\mathcal{R}h^{\mu\nu}a_\mu a_\nu\right]\right]\,.\label{eq:HLaction0}
\end{eqnarray}
The coefficients $c_1$ and $c_2$ have mass dimension $z$ and the coefficients $c_3$ and $c_4$ have mass dimension $2-z$. All the others are dimensionless. The terms with coefficients $c_3$ and $c_4$ are the kinetic terms because
\begin{equation}\label{eq:lambdaterm}
c_3 \nabla_\mu\hat v^\mu\nabla_\nu\hat v^\nu+c_4 \nabla_\nu\hat v^\mu\nabla_\mu\hat v^\nu=C\left(h^{\mu\rho}h^{\nu\sigma}K_{\mu\nu}K_{\rho\sigma}-\lambda\left(h^{\mu\nu}K_{\mu\nu}\right)^2\right)\,.
\end{equation}
The terms with coefficients $c_1$, $c_2$ and $c_{10}$ to $c_{15}$ only involve spatial derivatives and belong to the potential term $\mathcal{V}$. They agree with the potential terms in \cite{Blas:2009qj,Blas:2010hb,Zhu:2011yu} taking into consideration that we are in 2+1 dimensions. The terms with coefficients $c_5$ to $c_9$ involve mixed time and space derivatives and are in particular odd under time reversal. Hence in order to not to break time reversal invariance we will set these coefficients equal to zero. All other terms are time reversal and parity preserving. We thus obtain
\begin{equation}\label{eq:HLaction}
S = \int d^3x e\left[C\left(h^{\mu\rho}h^{\nu\sigma}K_{\mu\nu}K_{\rho\sigma}-\lambda\left(h^{\mu\nu}K_{\mu\nu}\right)^2\right)-\mathcal{V}\right]\,,
\end{equation}
where the potential $\mathcal{V}$ is given by
\begin{eqnarray}
\hspace{-.7cm}-\mathcal{V} & = & 2\Lambda+c_1 h^{\mu\nu}a_\mu a_\nu+c_2 \mathcal{R}+\delta_{z,2}\left[c_{10}\left(h^{\mu\nu}a_\mu a_\nu\right)^2+c_{11}h^{\mu\rho}a_\mu a_\rho\nabla_\nu\left(h^{\nu\sigma}a_\sigma\right)\right.\nonumber\\
\hspace{-1cm}&&\left.+c_{12}\nabla_\nu\left(h^{\mu\rho}a_\rho\right)\nabla_\mu\left(h^{\nu\sigma}a_\sigma\right)+c_{13}\mathcal{R}^2+c_{14}\mathcal{R}\nabla_\mu\left(h^{\mu\nu}a_\nu\right)+c_{15}\mathcal{R}h^{\mu\nu}a_\mu a_\nu\right]\,,\label{eq:potential}
\end{eqnarray}
which also includes a cosmological constant $\Lambda$. The kinetic terms in \eqref{eq:HLaction} display the $\lambda$ parameter of \cite{Horava:2008ih,Horava:2009uw}. The potential is exactly the same as the 3D version of the potential given in \cite{Blas:2009qj,Blas:2010hb,Zhu:2011yu}. We will not impose that $\mathcal{V}$ obeys the detailed balance condition. In the ADM parametrization of section \ref{sec:parametrization} the extrinsic curvature terms in \eqref{eq:lambdaterm} are just 
\begin{equation}
\gamma^{ik}\gamma^{jl}K_{ij}K_{kl}-\lambda\left(\gamma^{ij}K_{ij}\right)^2\,,
\end{equation}
where $K_{ij}$ is given by
\begin{equation}
K_{ij}=\frac{1}{2N}\left(\partial_t\gamma_{ij}-\mathcal{L}_N\gamma_{ij}\right)=\frac{1}{2N}\left(\partial_t\gamma_{ij}-\nabla^{(\gamma)}_i N_j-\nabla^{(\gamma)}_j N_i\right)\,,
\end{equation}
where $N_i=\gamma_{ij}N^j$ and $\nabla^{(\gamma)}_i$ is the covariant derivative that is metric compatible with respect to $\gamma_{ij}$.

\section{Local Bargmann Invariance of the HL Action: Local $U(1)$ vs St\"uckelberg Coupling}\label{sec:localU1}
 
The action \eqref{eq:HLaction} is by construction invariant under local Galilean transformations because it depends only on the invariants $\tau_\mu$ and $\hat h_{\mu\nu}$. So far we did not consider the possibility of adding $\tilde\Phi$. The action \eqref{eq:HLaction} is not invariant under the central extension of the Galilean algebra. We will now study what happens when we vary $m_\mu$ in \eqref{eq:HLaction} as $\delta m_\mu=\partial_\mu\sigma$. We have that the connection \eqref{eq:hatGammaTNC} transforms under the central element $N$ of the Bargmann algebra as
\begin{eqnarray}
\delta_N\Gamma^\rho_{\mu\nu} & = & \frac{1}{2}h^{\rho\lambda}\left[\left(a_\mu\tau_\nu-a_\nu\tau_\mu\right)\partial_\lambda\sigma+a_\lambda\tau_\nu\partial_\mu\sigma+a_\lambda\tau_\mu\partial_\nu\sigma\right]\nonumber\\
&&+h^{\rho\lambda}\tau_\mu\tau_\nu\left[\partial_\lambda\left(\hat v^\kappa\partial_\kappa\sigma\right)+2a_\lambda \hat v^\kappa\partial_\kappa\sigma\right]\,.
\end{eqnarray}
Using that $\Omega_\mu{}^a{}_b$ is given via \eqref{eq:covdere} and \eqref{eq:VP-2} by 
\begin{equation}
\Omega_\mu{}^a{}_b=e^\nu_b\left(\partial_\mu e_\nu^a-\Gamma^\rho_{\mu\nu}e_\rho^a\right)\,,
\end{equation}
we obtain
\begin{equation}
\delta_N\Omega_\mu{}^a{}_b=\frac{1}{2}\tau_\mu e^\nu_b e^{\lambda a}\left(a_\nu\partial_\lambda\sigma-a_\lambda\partial_\nu\sigma\right)\,.
\end{equation}
This implies that
\begin{equation}
\delta_N R_{abcd}(J)=0\,.
\end{equation}
Further $h^{\mu\nu}a_\nu$ is gauge invariant. Using the above results it can be shown that the whole potential $\mathcal{V}$ in \eqref{eq:potential} is gauge invariant. What is left is to transform the kinetic terms under $N$. We have
\begin{equation}
\delta_N\left(\nabla_\nu\hat v^\mu\nabla_\mu\hat v^\nu-\nabla_\mu\hat v^\mu\nabla_\nu\hat v^\nu\right)=-\mathcal{R}\hat v^\mu\partial_\mu\sigma+2\left(h^{\mu\lambda}a_\mu\nabla_\nu\hat v^\nu-h^{\mu\nu}a_\mu\nabla_\nu\hat v^\lambda\right)\partial_\lambda\sigma\,,
\end{equation}
where we used \eqref{eq:3DBI}. The first term can be cancelled by adding $\tilde\Phi\mathcal{R}$ to the action using that $\tilde\Phi$ transforms as
\begin{equation}
\delta_N\tilde\Phi=-\hat v^\mu\partial_\mu\sigma=N^{-1}\left(\partial_t\sigma-N^i\partial_i\sigma\right)\,,
\end{equation}
where in the second equality we expressed the results in terms of the ADM parameterization of  section \ref{sec:parametrization}. 

In \cite{Horava:2010zj} the following $U(1)$ transformation was introduced
\begin{equation}
\delta_\alpha N_i=N\partial_i\alpha\,.
\end{equation}
Together with two new fields $A$ and $\nu$ transforming as
\begin{eqnarray}
\delta_\alpha A & = & \partial_t\alpha-N^i\partial_i\alpha\,,\\
\delta_\alpha \nu & = & -\alpha\,,
\end{eqnarray} 
with $\nu$ called the Newtonian prepotential \cite{Horava:2010zj}. We see that the $\alpha$ transformation is none other than the Bargmann extension (the $\sigma$ transformation here) as follows from the identification of $m_i$ in \eqref{eq:mi}. More precisely we have $\alpha=-\sigma$. We thus see that the $A$ and $\nu$ fields can be identified with $\tilde\Phi$ and $\chi$ as follows: $A=-N\tilde\Phi$ and $\nu=\chi$. The term $\int d^3x e\mathcal{R}\tilde\Phi$ is what in \cite{Horava:2010zj} is denoted by $\int d^3x\sqrt{\gamma}A\mathcal{R}$. If we work in the context of projectable HL gravity for which $a_\mu=0$ the action \eqref{eq:HLaction} with $\lambda=1$ can be made $U(1)$ invariant by writing
\begin{equation}\label{eq:HLaction2}
S = \int d^3x e\left[C\left(h^{\mu\rho}h^{\nu\sigma}K_{\mu\nu}K_{\rho\sigma}-\left(h^{\mu\nu}K_{\mu\nu}\right)^2-\tilde\Phi\mathcal{R}\right)-\mathcal{V}\right]\,.
\end{equation}
However if we work with the non-projectable version or with $\lambda\neq 1$ we still need to add additional terms to make the theory $U(1)$ invariant. To see this we use the St\"uckelberg scalar $\chi$ that we already mentioned under \eqref{eq:tildePhi} (see also \cite{Hartong:2015wxa}). Using the field $\chi$ that transforms as $\delta\chi=\sigma$ we can construct the following gauge invariant action (the invariance is up to a total derivative) for $\lambda=1$
\begin{eqnarray}\label{eq:HLaction3}
S & = & \int d^3x e\left[C\left(h^{\mu\rho}h^{\nu\sigma}-h^{\mu\nu}h^{\rho\sigma}\right)\left(K_{\mu\nu}K_{\rho\sigma}-2a_\mu\partial_\nu\chi K_{\rho\sigma}+a_\rho\partial_\sigma\chi\nabla_\mu\partial_\nu\chi\right.\right.\nonumber\\
&&\left.\left.+\frac{1}{2}a_\mu a_\rho\partial_\nu\chi\partial_\sigma\chi\right)-C\tilde\Phi\mathcal{R}-\mathcal{V}\right]\,.
\end{eqnarray}
The $\chi$ dependent terms agree with the result of \cite{Zhu:2011yu} (eq. (3.8) of that paper)\footnote{To ease comparison it is useful to note that in the notation of \cite{Zhu:2011yu} one has the identity
\begin{eqnarray*}
&&\frac{1}{3}\hat{\mathcal{G}}^{ijkl}\left[4\left(\nabla_i\nabla_j\varphi\right)a_{(k}\nabla_{l)}\varphi+2\left(\nabla_{(i}\varphi\right)a_{j)(k}\nabla_{l)}\varphi+5a_{(i}\left(\nabla_{j)}\varphi\right)a_{(k}\nabla_{l)}\varphi\right]=\nonumber\\
&&\hat{\mathcal{G}}^{ijkl}\left[\left(\nabla_i\nabla_j\varphi\right)a_{k}\nabla_{l}\varphi+\frac{1}{2}a_j a_k\nabla_i\varphi\nabla_l\varphi\right]+\text{tot.der.}\,,
\end{eqnarray*}
where in the notation of \cite{Zhu:2011yu} the field $\varphi$ is what we call $\chi$. We also note that the coefficients of these terms are dimension independent.}. We thus see that when there is torsion $a_\mu\neq 0$ we need to introduce a St\"uckelberg scalar $\chi$ to make the action $U(1)$ invariant. While when there is no torsion we can use \eqref{eq:HLaction2}. This nicely agrees with the comments made below \eqref{eq:torsion}. In \cite{Horava:2010zj} the $\chi$ field is denoted by $\nu$. This means that we have the following invariance $\delta_N m_\mu=\partial_\mu\sigma$ and $\delta_N\chi=\sigma$. As a consequence we may simply replace everywhere $m_\mu$ by $M_\mu=m_\mu-\partial_\mu\chi$. This is consistent with the observations made in \cite{daSilva:2010bm} (see in particular eq. (20) of said paper). Essentially adding the $\chi$ field to the action means that we have trivialized the $U(1)$ symmetry by St\"uckelberging it or in other words we have removed the $U(1)$ transformations all together (see the next section). 

Let us define $K_{\mu\nu}^\chi$ as \eqref{eq:extrinsiccurv} with $m_\mu$ replaced by $M_\mu$. It can be shown that
\begin{equation}\label{eq:U(1)invK}
h^{\mu\rho}h^{\nu\sigma}K^\chi_{\mu\nu}=h^{\mu\rho}h^{\nu\sigma}\left(K_{\rho\sigma}-\nabla_\rho\partial_\sigma\chi-\frac{1}{2}a_\rho\partial_\sigma\chi-\frac{1}{2}a_\sigma\partial_\rho\chi\right)\,,
\end{equation}
which is now by construction manifestly $U(1)$ invariant. Similarly we can write a manifestly $U(1)$ invariant $\tilde\Phi$ as
\begin{equation}\label{eq:U(1)invtildePhi}
\tilde\Phi_\chi=\tilde\Phi+\hat v^\mu\partial_\mu\chi+\frac{1}{2}h^{\mu\nu}\partial_\mu\chi\partial_\nu\chi\,,
\end{equation}
obtained by replacing $m_\mu$ by $M_\mu$ in $\tilde\Phi$. Instead of \eqref{eq:HLaction3} we then write 
\begin{equation}\label{eq:HLaction4}
S = \int d^3x e\left[C\left(h^{\mu\rho}h^{\nu\sigma}K^\chi_{\mu\nu}K^\chi_{\rho\sigma}-\left(h^{\mu\nu}K^\chi_{\mu\nu}\right)^2-\tilde\Phi_\chi\mathcal{R}\right)-\mathcal{V}\right]\,.
\end{equation}
It can be checked that this is up to total derivative terms the same as \eqref{eq:HLaction3}. It is now straightforward to generalize this to arbitrary $\lambda$ and to add for example the $\Omega\tilde\Phi$ coupling considered in \cite{Horava:2010zj} leading to 
\begin{equation}\label{eq:HLaction5}
S = \int d^3x e\left[C\left(h^{\mu\rho}h^{\nu\sigma}K^\chi_{\mu\nu}K^\chi_{\rho\sigma}-\lambda\left(h^{\mu\nu}K^\chi_{\mu\nu}\right)^2-\tilde\Phi_\chi\left(\mathcal{R}-2\Omega\right)\right)-\mathcal{V}\right]\,.
\end{equation}
If we isolate the part of the action that depends on $\chi$ we find precisely the same answer as in eq. (3.12) of \cite{Zhu:2011yu} specialized to 2+1 dimensions\footnote{This simply means that we can take in the notation of  \cite{Zhu:2011yu} $G^{ij}=0$.}.

As a final confirmation that TNC geometry is a natural framework for HL gravity we will show 
in Section \ref{sec:confHLactions} that the conformal HL gravity theories can be obtained by adding dilatations to the Bargmann algebra, i.e. by considering the Schr\"odinger algebra.

\section{A Constraint Equation}\label{sec:constraint}

What we have learned is that unless the $\chi$ field drops out of the action, as in \eqref{eq:HLaction2} for the case of projectable HL gravity with $\lambda=1$, we no longer have a non-trivial local $U(1)$ invariance. This is because we can express everything in terms of $M_\mu$ which is inert under the $U(1)$. Essentially the fact that we had to introduce a St\"uckelberg scalar tells us that the $U(1)$ was not there in the first place. 

There are several statements in the literature expressing that one can remove a scalar degree of freedom from the theory by employing the $U(1)$ invariance, but since we have just established that unless we are dealing with \eqref{eq:HLaction2} there is no $U(1)$ these statements are not clear to us. What we will show is that there is a different mechanism that essentially accomplishes the same effect, via a constraint equation obtained by varying $\tilde\Phi_\chi$ in \eqref{eq:HLaction5}, to the claims made in the literature.

Since $\tilde\Phi_\chi$ is a field like any other we should, in order to be fully general, allow for arbitrary couplings to $\tilde\Phi_\chi$ that do not lead to terms of dimension higher than $z+2$. Put another way the most general HL action can be obtained by writing down the most general action depending on $\tau_\mu$, $\hat h_{\mu\nu}$ and $\tilde\Phi_\chi$ containing terms up to order (dilatation weight) $z+2$. The first thing to notice is that we typically cannot write down a kinetic term for $\tilde\Phi_\chi$ because the dilatation weight of $\left(\hat v^\mu\partial_\mu\tilde\Phi_\chi\right)^2$ is $6z-4$ which is larger than $z+2$ whenever $z>6/5$. The same is true for $K\hat v^\mu\partial_\mu\tilde\Phi_\chi$ while a term like $\hat v^\mu\partial_\mu\tilde\Phi_\chi$ or what is the same upon partial integration $K\tilde\Phi_\chi$ breaks time reversal invariance. Let us assume that we have a $z$ value larger than $6/5$ so that we cannot write a kinetic term. This means that $\tilde\Phi_\chi$ will appear as a non-propagating scalar field. 

Let us enumerate the possible allowed couplings to $\tilde\Phi_\chi$. Starting with the kinetic terms we can have schematically $\tilde\Phi_\chi^\alpha K^2$ where by $K^2$ we mean both ways of contracting the product of two extrinsic curvatures. In order for this term to have a dimension less than or equal to $z+2$ we need that $\alpha\le \frac{2-z}{2(z-1)}$. It follows that for $z>4/3$ we need $\alpha<1$. Consider next a term of the form $\tilde\Phi_\chi^\beta X$ where $X$ is any term of dimension 2. The condition that the weight does not exceed $z+2$ gives us $\beta\le \frac{z}{2(z-1)}$ which means that if $z>4/3$ we need $\beta<2$. Finally we can have terms of the form $\tilde\Phi_\chi^\gamma$ where $\gamma\le\frac{2+z}{2(z-1)}$ so that for $z>8/5$ we need that $\gamma<3$. In particular it is allowed for all values of $1<z\le 2$ to add a term of the form $\tilde\Phi_\chi^2$.

Since for $z>6/5$ we are not allowed to add a kinetic term for $\tilde\Phi_\chi$ we can integrate it out. We demand that the resulting action after integrating out $\tilde\Phi_\chi$ is local. This puts constraints on what $\alpha$, $\beta$ and $\gamma$ can be since they influence the solution for $\tilde\Phi_\chi$. We assume here that $\alpha$, $\beta$ and $\gamma$ are non-negative integers. We will be interested in values of $z$ close to $z=2$ so we assume that $z>8/5$. In that case we have the following allowed non-negative integer values: $\alpha=0$, $\beta=0,1$ and $\gamma=0,1,2$. In other words we can add the following $\tilde\Phi_\chi$ dependent terms
\begin{equation}
\tilde\Phi_\chi\left(d_1+d_2\mathcal{R}+d_3\nabla_\mu\left(h^{\mu\nu}a_\nu\right)+d_4h^{\mu\nu}a_\mu a_\nu+d_5\tilde\Phi_\chi\right)\,.
\end{equation}

There are now two cases of interest: either $d_5\neq 0$ or $d_5=0$. When $d_5\neq 0$ we can solve for $\tilde\Phi_\chi$ and substitute the result back into the action. The resulting action will be of the same form as \eqref{eq:HLaction} where all the terms originating from solving for $\tilde\Phi_\chi$ and substituting the result back into the action can be absorbed into the potential terms by renaming the coefficients in $\mathcal{V}$. The other possibility that $d_5=0$ leads to a rather different situation. In that case the equation of motion of $\tilde\Phi_\chi$ leads to the constraint equation 
\begin{equation}\label{eq:constraint}
d_1+d_2\mathcal{R}+d_3\nabla_\mu\left(h^{\mu\nu}a_\nu\right)+d_4h^{\mu\nu}a_\mu a_\nu=0\,.
\end{equation}
The remaining equations of motion for $\tau_\mu$ etc. will depend on $\tilde\Phi_\chi$ because there is no local symmetry (in particular no $U(1)$) that allows us to gauge fix this field to zero. Since there is no kinetic term for $\tilde\Phi_\chi$, and hence its value will not be determined dynamically, we fix it by means of a Lagrange multiplier term. Recall that for any value of $z$ in the range $1<z\le 2$ it is allowed by the effective action approach to add a term proportional to $\tilde\Phi_\chi^2$. Consider now the following action
\begin{equation}\label{eq:HLaction6}
S = \int d^3x e\left[\text{$\tilde\Phi_\chi$ indep. part}+\tilde\Phi_\chi\left(d_1+d_2\mathcal{R}+d_3\nabla_\mu\left(h^{\mu\nu}a_\nu\right)+d_4h^{\mu\nu}a_\mu a_\nu\right)+\lambda\tilde\Phi_\chi^2\right]\,,
\end{equation}
where crucially now $\lambda$ is a field, i.e. a Lagrange multiplier, so that its equation of motion tells us that $\tilde\Phi_\chi=0$ and further the equation of motion of $\tilde\Phi_\chi$ will lead to the constraint equation \eqref{eq:constraint}, which is a more general version of the constraint equation used in \cite{Horava:2010zj} and related works. Since $\tilde\Phi_\chi=0$ the $\tilde\Phi_\chi$ dependent terms do not affect the remaining equations of motion. This essentially accomplishes that $\tilde\Phi_\chi$ is not present in the theory and that we have the constraint equation \eqref{eq:constraint}. More generally we should think of $\tilde\Phi_\chi$ as a background field whose value can be set to be equal to some fixed function $f$. This is accomplished by writing instead of \eqref{eq:HLaction6} the following
\begin{eqnarray}
S & = &  \int d^3x e\left[\text{$\tilde\Phi_\chi$ indep. part}+\left(\tilde\Phi_\chi-f\right)\left(d_1+d_2\mathcal{R}+d_3\nabla_\mu\left(h^{\mu\nu}a_\nu\right)+d_4h^{\mu\nu}a_\mu a_\nu\right)\right.\nonumber\\
&&\left.+\lambda\left(\tilde\Phi_\chi-f\right)^2\right]\,.\label{eq:HLaction7}
\end{eqnarray}
The $\lambda$ equation of motion enforces the background value $\tilde\Phi_\chi=f$, the equation of motion of $\tilde\Phi_\chi$ leads again to \eqref{eq:constraint} while the remaining equations of motion involve terms depending on $f$ through the variation of terms linear in $f$.

\section{Conformal HL Gravity from the Schr\"odinger Algebra}\label{sec:confHLactions}

In this section we will work with an arbitrary number of spatial dimensions $d$. In order to study conformal HL actions we add dilatations to the Bargmann algebra of section \ref{sec:Bargmann} and study the various conformal invariants that one can build. To this end we use the connection $\mathcal{A}_\mu$ that takes values in the Schr\"odinger algebra (where for $z=2$ we leave out for now the special conformal transformations that will be introduced later)\footnote{Compared to e.g. \cite{Hartong:2014oma,Bergshoeff:2014uea} we have interchanged the field $m_\mu$ appearing in front of $N$ in the Bargmann algebra and the field $\tilde m_\mu$ appearing in front of $N$ in the Schr\"odinger algebra, see also footnote \ref{fn:notation}.}
\begin{equation}\label{eq:YMcurvSch}
\mathcal{A}_\mu = H\tau_{\mu}+P_ae_{\mu}^a+G_a\omega_{\mu}{}^a+\frac{1}{2}J_{ab}\omega_{\mu}{}^{ab}+N\tilde m_\mu+D b_\mu\,,
\end{equation}
where the new connection $b_\mu$ is called the dilatation connection. The reason that we renamed the connections in \eqref{eq:YMcurvSch} as compared to \eqref{eq:curlABargmann} is because the dilatation generator $D$ is not central so that it modifies the transformations under local $D$ transformations as compared to how say $\Omega_{\mu}{}^a$ and $\omega_{\mu}{}^{ab}$ would transform using \eqref{eq:covdere}, \eqref{eq:VP-2} and \eqref{eq:GammaTNC}. The transformation properties and curvatures of the various fields follow from the Schr\"odinger algebra:
\begin{equation}\label{eq:Schalgebra}
\begin{array}{ll}
\left[D\,,H\right] = -zH\,,   &  \left[D\,,P_a\right] = -P_a\,,\\
\left[D\,,G_a\right] = (z-1)G_a\,,   &  \left[D\,,N\right] = (z-2)N\,,\\
\left[H\,,G_a\right] = P_a\,, & \left[P_a\,,G_b\right] = \delta_{ab}N\,,\\
\left[J_{ab}\,,P_c\right] = \delta_{ac}P_b-\delta_{bc}P_a\,, &
\left[J_{ab}\,,G_c\right] = \delta_{ac}G_b-\delta_{bc}G_a\,,\\
\left[J_{ab}\,,J_{cd}\right] = \delta_{ac}J_{bd}-\delta_{ad}J_{bc}-\delta_{bc}J_{ad}+\delta_{bd}J_{ac}\,. &
\end{array}
\end{equation}
We perform the same steps as before (see \eqref{eq:YMtrafo} and onwards), namely we consider the adjoint transformation of $\mathcal{A}_\mu$, i.e.
\begin{equation}\label{eq:deltaASch}
\delta{\mathcal{A}}_{\mu}=\partial_\mu\Lambda+[{\mathcal{A}}_{\mu}\,,\Lambda]\,,
\end{equation}
where we write (without loss of generality)
\begin{equation}
\Lambda=\xi^\mu\mathcal{A}_\mu+\Sigma\,,
\end{equation}
with now
\begin{equation}
\Sigma = G_a\lambda^a+\frac{1}{2}J_{ab}\lambda^{ab}+N\sigma+D\Lambda_D\,,
\end{equation}
and we define $\bar\delta\mathcal{A}_\mu$ as
\begin{equation}\label{eq:bardeltaAagain}
\bar\delta\mathcal{A}_\mu=\delta\mathcal{A}_\mu-\xi^\nu\mathcal{F}_{\mu\nu}=\mathcal{L}_\xi\mathcal{A}_\mu+\partial_\mu\Sigma+[{\mathcal{A}}_{\mu}\,,\Sigma]\,,
\end{equation}
where $\mathcal{F}_{\mu\nu}$ is the curvature
\begin{eqnarray}
\mathcal{F}_{\mu\nu} & = & \partial_\mu\mathcal{A}_\nu-\partial_\nu\mathcal{A}_\mu+[{\mathcal{A}}_{\mu}\,,{\mathcal{A}}_{\nu}]\nonumber\\
&=& H\tilde R_{\mu\nu}(H)+P_a\tilde R_{\mu\nu}{}^a(P)+G_a\tilde R_{\mu\nu}{}^a(G)+\frac{1}{2}J_{ab}\tilde R_{\mu\nu}{}^{ab}(J)+N\tilde R_{\mu\nu}(N)\nonumber\\
&&+D\tilde R_{\mu\nu}(D)\,,
\end{eqnarray}
where we put tildes on the curvatures to distinguish them from those given in sections \ref{sec:LocalGalTrafos} and \ref{sec:Bargmann}. From this we obtain among others that the dilatation connection $b_\mu$ transforms as 
\begin{equation}\label{eq:trafob}
\bar\delta b_\mu = \mathcal{L}_\xi b_\mu+\partial_\mu\Lambda_D\,.
\end{equation}
The following discussion closely follows section 4 of \cite{Bergshoeff:2014uea}. We will use this $b_\mu$ connection to rewrite the covariant derivatives \eqref{eq:covdertau} and \eqref{eq:covdere} in a manifestly dilatation covariant manner. 

As a note on our notation we remark that, now that we have learned that we should work with $M_\mu=m_\mu-\partial_\mu\chi$ we take it for granted that we have replaced everywhere $m_\mu$ by $M_\mu$ and we from now on suppress $\chi$ labels as in \eqref{eq:U(1)invK} and \eqref{eq:U(1)invtildePhi}. The Schr\"odinger algebra for general $z$ tells us that the dilatation weights of the fields are as in table \ref{table:dimensionsinvariants} while $m_\mu$ and $\chi$ (and thus $M_\mu$) have dilatation weight $z-2$. This also agrees with the weights assigned to $A$ and $\nu$ in \cite{Horava:2010zj}.

Coming back to the introduction of $b_\mu$, to make expressions dilatation covariant we take $\bar\Gamma^\rho_{\mu\nu}$ of equation \eqref{eq:barGammaTNC} and replace ordinary derivatives by dilatation covariant ones leading to a new connection $\tilde\Gamma^{\rho}_{\mu\nu}$ that is invariant under the $G_a$, $J_{ab}$, $N$ and $D$ transformations and which is given by \cite{Bergshoeff:2014uea}
\begin{equation}\label{eq:tildeGammaTNC}
\tilde\Gamma^{\rho}_{\mu\nu} = -\hat v^\rho\left(\partial_\mu-zb_\mu\right)\tau_\nu+\frac{1}{2}h^{\rho\sigma}\left(\left(\partial_\mu-2b_\mu\right)\bar h_{\nu\sigma}+\left(\partial_\nu-2b_\nu\right) \bar h_{\mu\sigma}-\left(\partial_\sigma-2b_\sigma\right)\bar h_{\mu\nu}\right)\,.
\end{equation}
For the most part of this section we will work with $\bar\Gamma^\rho_{\mu\nu}$ and its dilatation covariant generalization $\tilde\Gamma^\rho_{\mu\nu}$. The final scalars out of which we will build the HL action, i.e. for dynamical TTNC geometries, are such that it does not matter whether we use $\bar \Gamma^\rho_{\mu\nu}$ or $\hat\Gamma^\rho_{\mu\nu}$ which are related via \eqref{eq:relationGammahatbar}.

With the help of $b_\mu$ and $\tilde\Gamma^{\rho}_{\mu\nu}$ we can now rewrite the covariant derivatives \eqref{eq:covdertau} and \eqref{eq:covdere} as follows
\begin{eqnarray}
\mathcal{D}_\mu\tau_\nu & = &\partial_\mu\tau_\nu-\tilde\Gamma^{\rho}_{\mu\nu}\tau_\rho-zb_\mu\tau_\nu=0\,,\label{eq:covder1v2}  \\
\mathcal{D}_\mu e_\nu{}^a & = & \partial_\mu e_\nu{}^a-\tilde\Gamma^{\rho}_{\mu\nu}e_\rho{}^a-\omega_\mu{}^a\tau_\nu-\omega_\mu{}^{a}{}_be_{\nu}{}^b-b_\mu e_\nu{}^a=0 \,. \label{eq:covder2v2}
\end{eqnarray}
The $\omega_\mu{}^a$ and $\omega_\mu{}^{ab}$ connections are such that they can be written in terms of $\Omega_\mu{}^a$ and $\Omega_\mu{}^{ab}$ together with $b_\mu$ dependent terms such that all the $b_\mu$ terms drop out on the right hand side of \eqref{eq:covder1v2} and \eqref{eq:covder2v2} when expressing it in terms of the connections $\Gamma^{\rho}_{\mu\nu}$, $\Omega_\mu{}^a$ and $\Omega_\mu{}^{ab}$.

The field $M_\mu=m_\mu-\partial_\mu\chi$ can be expressed in terms of the Schr\"odinger connection $\tilde m_\mu$ as follows. According to \eqref{eq:Schalgebra} and \eqref{eq:bardeltaAagain} the Schr\"odinger connection $\tilde m_\mu$ transforms as
\begin{equation}
\bar\delta \tilde m_\mu=\mathcal{L}_\xi\tilde m_\mu+\partial_\mu\sigma+\lambda^a e_{\mu a}+(z-2)\left(\sigma b_\mu-\Lambda_D\tilde m_\mu\right)\,.
\end{equation}
The St\"uckelberg scalar $\chi$ transforms as
\begin{equation}
\bar\delta\chi=\mathcal{L}_\xi\chi+\sigma+(2-z)\Lambda_D\chi\,.
\end{equation}
A Schr\"odinger covariant derivative $\mathcal{D}_\mu\chi$ is given by
\begin{equation}
\mathcal{D}_\mu\chi=\partial_\mu\chi-\tilde m_\mu-(2-z)b_\mu\chi\,.
\end{equation}
Defining $M_\mu=-\mathcal{D}_\mu\chi=m_\mu-\partial_\mu\chi$ we see that $M_\mu$ transforms as
\begin{equation}\label{eq:trafoM}
\bar\delta M_\mu=\mathcal{L}_\xi M_\mu+e_\mu{}^a\lambda_a+(2-z)\Lambda_D M_\mu\,,
\end{equation}
and that 
\begin{equation}
m_\mu=\tilde m_\mu+(2-z)b_\mu\chi\,.
\end{equation}
Hence the dilatation covariant derivative of $M_\mu$ reads
\begin{equation}
\mathcal{D}_\mu M_\nu=\partial_\mu M_\nu-\tilde\Gamma^{\rho}_{\mu\nu}M_\rho-(2-z)b_\mu M_\nu-\omega_\mu{}^a e_{\nu a}\,.
\end{equation}

The torsion $\tilde\Gamma^{\rho}_{[\mu\nu]}$ has to be a $G$, $J$, $N$ and $D$ invariant tensor. With our TTNC field content the only option is to take it zero, i.e. $\tilde\Gamma^\rho_{\mu\nu}$ becomes torsionless \cite{Bergshoeff:2014uea}. This means that different from the relativistic case the $b_\mu$ connection is not entirely independent, but instead reads 
\begin{equation}\label{eq:TNCb}
b_\mu=\frac{1}{z}\hat v^\rho\left(\partial_\rho\tau_\mu-\partial_\mu\tau_\rho\right)-\hat v^\rho b_\rho \tau_\mu=\frac{1}{z}a_\mu-\hat v^\rho b_\rho \tau_\mu\,.
\end{equation}

Let $X^\rho$ be a tensor with dilatation weight $w$, i.e.
\begin{equation}
\delta_D X^\rho=-w\Lambda_D X^\rho\,.
\end{equation}
A dilatation covariant derivative is given by
\begin{equation}
\tilde\nabla_\nu X^\rho+wb_\nu X^\rho\,,
\end{equation}
where $\tilde\nabla_\nu$ is covariant with respect to $\tilde\Gamma^\rho_{\nu\mu}$ as given in \eqref{eq:tildeGammaTNC}. Let us compute the commutator
\begin{eqnarray}
&&\left(\tilde\nabla_\mu+wb_\mu\right)\left(\tilde\nabla_\nu+wb_\nu\right)X^\rho-(\mu\leftrightarrow\nu)\nonumber\\
&=&-\tilde R_{\mu\nu\lambda}{}^\rho X^\lambda+w\left(\partial_\mu b_\nu-\partial_\nu b_\mu\right)X^\rho\,,
\end{eqnarray}
where 
\begin{equation}
\tilde R_{\mu\nu\lambda}{}^\rho=-\partial_\mu\tilde\Gamma^\rho_{\nu\lambda}+\partial_\nu\tilde\Gamma^\rho_{\mu\lambda}-\tilde\Gamma^\rho_{\mu\sigma}\tilde\Gamma^\sigma_{\nu\lambda}+\tilde\Gamma^\rho_{\nu\sigma}\tilde\Gamma^\sigma_{\mu\lambda}\,.
\end{equation}

The introduction of the $b_\mu$ connection has led to a new component $\hat v^\mu b_\mu$ as visible in \eqref{eq:TNCb}. We can introduce a special conformal transformation (denoted by $K$) that allows us to remove this component. Hence we assign a new transformation rule to $b_\mu$ namely
\begin{equation}
\delta_K b_\mu=\Lambda_K\tau_\mu\,.
\end{equation}

Under special conformal transformations we have
\begin{equation}
\delta_K\tilde\Gamma^\rho_{\mu\nu}=\Lambda_K\left((z-2)\hat v^\rho\tau_\mu\tau_\nu-\delta^\rho_\mu\tau_\nu-\delta^\rho_\nu\tau_\mu\right)\,.
\end{equation}
In order that $\left(\tilde\nabla_\mu+wb_\mu\right)\left(\tilde\nabla_\nu+wb_\nu\right)X^\rho$ transforms covariantly we define the $K$-covariant derivative
\begin{eqnarray}
&&\left(\tilde{\mathcal{D}}_\mu+wb_\mu\right)\left(\tilde\nabla_\nu+wb_\nu\right)X^\rho=\left(\tilde\nabla_\mu+wb_\mu\right)\left(\tilde\nabla_\nu+wb_\nu\right)X^\rho\nonumber\\
&&-wf_\mu\tau_\nu X^\rho-f_\mu\left((z-2)\hat v^\rho\tau_\nu\tau_\lambda-\delta^\rho_\nu\tau_\lambda-\delta^\rho_\lambda\tau_\nu\right)X^\lambda\,,\label{eq:Kcovder}
\end{eqnarray}
where $f_\mu$ is a connection for local $K$ transformations that transforms as \cite{Bergshoeff:2014uea}
\begin{equation}\label{eq:trafo-f}
\bar\delta f_\mu=\mathcal{L}_\xi f_\mu+\partial_\mu\Lambda_K-z\Lambda_D f_\mu+z\Lambda_K b_\mu\,.
\end{equation}

In order not to introduce yet another independent field $f_\mu$ (recall that we are trying to remove $\hat v^\mu b_\mu$) we demand that $f_\mu$ is a completely dependent connection that transforms as in \eqref{eq:trafo-f}. This is in part realized by setting the curvature of the dilatation connection $b_\mu$ equal to zero, i.e. by imposing
\begin{equation}
\check R_{\mu\nu}(D)=\partial_\mu b_\nu-\partial_\nu b_\mu-f_\mu\tau_\nu+f_\nu\tau_\mu=0\,.
\end{equation}
This fixes all but the $\hat v^\mu f_\mu$ component of $f_\mu$. This latter component will be fixed later by equation \eqref{eq:fixingvf}. The notation is such that a tilde refers to a curvature of the $\bar\delta$ transformation \eqref{eq:bardeltaAagain} without the $K$ transformation while a curvature with a check sign refers to a curvature that is covariant under the $\bar\delta$ transformations with the $K$ transformation. We note that while for the Schr\"odinger algebra, i.e. with the $\delta$ transformations \eqref{eq:deltaASch} we can only add special conformal transformations when $z=2$ while for the (different) group of transformations transforming under $\bar\delta$ we can define $K$ transformations for any $z$ \cite{Bergshoeff:2014uea}.

Taking the commutator of \eqref{eq:Kcovder} we find 
\begin{equation}
\left(\tilde{\mathcal{D}}_\mu+wb_\mu\right)\left(\tilde\nabla_\nu+wb_\nu\right)X^\rho-(\mu\leftrightarrow\nu)=-\check R_{\mu\nu\lambda}{}^\rho X^\lambda\,,
\end{equation}
where $\check R_{\mu\nu\lambda}{}^\rho$ is given by
\begin{eqnarray}
\check R_{\mu\nu\lambda}{}^\rho & = & \tilde R_{\mu\nu\lambda}{}^\rho+(z-2)\hat v^\rho\tau_\lambda\left(f_\mu\tau_\nu-f_\nu\tau_\mu\right)-\delta^\rho_\nu\tau_\lambda f_\mu+\delta^\rho_\mu\tau_\lambda f_\nu\nonumber\\
&&-\delta^\rho_\lambda\left(f_\mu\tau_\nu-f_\nu\tau_\mu\right)\,.
\end{eqnarray}
Under $K$ transformations the curvature tensor $\check R_{\mu\nu\lambda}{}^\rho$ transforms as
\begin{equation}
\delta_K \check R_{\mu\nu\lambda}{}^\rho = \Lambda_K\left[-(z-2)\tau_\lambda\tau_\nu\mathcal{D}_\mu\hat v^\rho+(z-2)\tau_\lambda\tau_\mu\mathcal{D}_\nu\hat v^\rho\right]\,.\label{eq:KtrafobarRiemann}
\end{equation}
Besides this property, the tensor $\check R_{\mu\nu\lambda}{}^\rho$ is by construction invariant under $D$, $G$, $N$ and $J$ transformations.

Using the vielbein postulates \eqref{eq:covder1v2} and \eqref{eq:covder2v2} we can write
\begin{equation}
\tilde\Gamma^\rho_{\mu\nu}=-v^\rho\left(\partial_\mu\tau_\nu-zb_\mu\tau_\nu\right)+e^\rho_a\left(\partial_\mu e_\nu^a-\omega_\mu{}^a\tau_\nu-\omega_\mu{}^a{}_b e^b_\nu-b_\mu e^a_\nu\right)\,.
\end{equation}
With this result we can derive
\begin{equation}\label{eq:barRiemann}
\check R_{\mu\nu\sigma}{}^\rho = -e^{\rho d}e^c_\sigma\tilde R_{\mu\nu cd}(J)+e^{\rho c}\tau_\sigma\check R_{\mu\nu c}(G)\,,
\end{equation}
where $\tilde R_{\mu\nu cd}(J)$ and $\check R_{\mu\nu c}(G)$ are given by
\begin{eqnarray}
\tilde R_{\mu\nu}{}^{ab}(J) & = & 2\partial_{[\mu} \omega_{\nu]}{}^{ab}-2\omega_{[\mu}{}^{ca}\omega_{\nu]}{}^b{}_c\,,\label{eq:omegacurv}\\
\check R_{\mu\nu}{}^a(G) & = & \tilde R_{\mu\nu}{}^a(G)-2f_{[\mu}\left(e_{\nu]}{}^a+(z-2)\tau_{\nu]}M^a\right)\nonumber\\
&=&2\partial_{[\mu}\omega_{\nu]}{}^a-2\omega_{[\mu}{}^{ab}\omega_{\nu]b}-2(1-z)b_{[\mu}\omega_{\nu]}{}^a\nonumber\\
&&-2f_{[\mu}\left(e_{\nu]}{}^a+(z-2)\tau_{\nu]}M^a\right)\,.\label{eq:newGcurv}
\end{eqnarray}
We next present some basic properties of $\check R_{\mu\nu\sigma}{}^\rho$. The first is
\begin{equation}\label{eq:traceRiemann}
\check R_{\mu\nu\rho}{}^\rho=0\,,
\end{equation}
and the second is
\begin{equation}\label{eq:BI}
\check R_{[\mu\nu\sigma]}{}^\rho=0\,.
\end{equation}
Equations \eqref{eq:barRiemann} and \eqref{eq:BI} together give us the Bianchi identity 
\begin{equation}\label{eq:BInew}
\tilde R_{[\mu\nu}{}^{ab}(J)e_{\rho]b}+\check R_{[\mu\nu}{}^a(G)\tau_{\rho]}=0\,.
\end{equation}
By contracting this with $v^\mu e^\nu{}_c e^\rho{}_a$ we find
\begin{equation}\label{eq:BI1}
\check R_{ca}{}^a(G) + v^\mu \tilde R_{\mu a}{}^a{}_c(J) =0\,,
\end{equation}
and by contracting \eqref{eq:BInew} with $e^\mu{}_b e^\nu{}_a e^\rho{}_c$ we obtain
\begin{equation}\label{eq:BI2}
\tilde R_{ba}{}^a{}_c(J)-\tilde R_{ca}{}^a{}_b(J) =0\,.
\end{equation}
The two identities \eqref{eq:BI} and \eqref{eq:traceRiemann} imply that
\begin{equation}
\check R_{\rho[\nu\sigma]}{}^\rho=0\,.
\end{equation}
We define $\check R_{\nu\sigma}=\check R_{\nu\rho\sigma}{}^\rho$.

Using the identity \eqref{eq:BI1} we can derive
\begin{eqnarray}
\hspace{-1cm}\hat v^{\sigma}\hat v^\nu \check R_{\sigma\nu} & = & -\hat v^\nu\left(R_{\nu a}{}^a(G)+M^cR_{\nu a}{}^a{}_c\right)\nonumber\\
\hspace{-1cm}&=& -\hat v^\mu\left( R_{\mu a}{}^a(G)+2 M^c R_{\mu a}{}^a{}_c(J)\right)-M^b\left(R_{ba}{}^a(G)+M^c R_{ba}{}^a{}_c(J)\right)\,.
\end{eqnarray}
We now turn to the question what $\hat v^\sigma \hat v^\nu \check R_{\sigma\nu}$ should be equal to. Following \cite{Bergshoeff:2014uea} we will take this to be equal to
\begin{equation}\label{eq:fixingvf}
\hat v^{\sigma}\hat v^\nu \check R_{\sigma\nu}=\frac{1}{2d}(z-2)\left(h^{\mu\nu}\mathcal{D}_\mu M_\nu\right)^2\,,
\end{equation}
because the right hand side has the exact same transformation properties under all local symmetries as $\hat v^{\sigma}\hat v^\nu \check R_{\sigma\nu}$. The combination of $\check R_{\mu\nu}(D)=0$ together with \eqref{eq:fixingvf} fixes $f_\mu$ entirely in terms of $\tau_\mu$, $e_\mu^a$, $m_\mu$ and $\chi$ in such a way that it transforms as in \eqref{eq:trafo-f}.

Using that
\begin{equation}
h^{\mu\rho}h^{\nu\sigma}\mathcal{D}_{(\mu}M_{\nu)}=h^{\mu\rho}h^{\nu\sigma}\left(K_{\mu\nu}+\hat v^\lambda b_\lambda\hat h_{\mu\nu}\right)\,,
\end{equation}
where $K_{\mu\nu}$ is the extrinsic curvature, we see that
\begin{equation}
h^{\mu\rho}h^{\nu\sigma}\mathcal{D}_{(\mu}M_{\nu)}\mathcal{D}_{(\rho}M_{\sigma)}-\frac{1}{d}\left(h^{\mu\nu}\mathcal{D}_\mu M_\nu\right)^2\,,
\end{equation}
is invariant under the $K$ transformation because the term $\hat v^\mu b_\mu$ cancels out from the above difference. Another scalar quantity of interest is 
\begin{equation}
h^{\mu\nu}\check R_{\mu\nu}=-\tilde R^{ab}{}_{ab}(J)\,,
\end{equation}
which is $K$ invariant and has dilatation weight $2$. With these ingredients we can build a $z=d$ conformally invariant Lagrangian 
\begin{equation}
\mathcal{L} = e\left[A\left(h^{\mu\rho}h^{\nu\sigma}K_{\mu\nu}K_{\rho\sigma}-\frac{1}{d}\left(h^{\mu\nu}K_{\mu\nu}\right)^2\right) +B\left(h^{\mu\nu}\check R_{\mu\nu}\right)^d\right]\,.
\end{equation}
This is an example of a Lagrangian for non-projectable HL gravity that is conformally invariant. 

The quantity $h^{\mu\nu}\check R_{\mu\nu}$ can be expressed in terms of $\mathcal{R}$ and the torsion vector $a_\mu$ defined in sections \ref{sec:torsion} and \ref{sec:curvatures} as follows. Solving \eqref{eq:covder2v2} for $\omega_\mu{}^{ab}$ and using the relation between $\tilde\Gamma^\rho_{\mu\nu}$ and $\bar \Gamma^\rho_{\mu\nu}$ given in \eqref{eq:barGammaTNC} which reads
\begin{equation}
\tilde\Gamma^\rho_{\mu\nu}=\bar\Gamma^\rho_{\mu\nu}+z\hat v^\rho b_\mu\tau_\nu-h^{\rho\sigma}\left(b_\mu\bar h_{\nu\sigma}+b_\nu\bar h_{\mu\sigma}-b_\sigma \bar h_{\mu\nu}\right)\,,
\end{equation}
we obtain
\eqref{eq:covder2v2}, via
\begin{equation}\label{eq:Omega-vs-omega}
\omega_{\mu}{}^{ab}=\hat\Omega_\mu{}^{ab}+e^{\nu b}b_\nu\hat e_{\mu}^a-e^{\nu a}b_\nu\hat e_{\mu}^b\,,
\end{equation}
where we used that $\omega_\mu{}^{ab}$ and $\Omega_\mu{}^{ab}$ are related, as follows from the vielbein postulates \eqref{eq:covdere}, \eqref{eq:VP-2} and where we furthermore used that for TTNC $\bar\Omega_{\mu}{}^{ab}=\hat \Omega_\mu{}^{ab}$ as follows from \eqref{eq:relationGammahatbar} and the TTNC relation \eqref{eq:TTNCidentity1}. In the relation $\bar\Omega_{\mu}{}^{ab}=\hat \Omega_\mu{}^{ab}$ the connection $\bar\Omega_{\mu}{}^{ab}$ is found by employing the vielbein postulate expressed in terms of $\bar\Gamma^\rho_{\mu\nu}$ and likewise $\hat\Omega_{\mu}{}^{ab}$ is obtained by using the vielbein postulate written in terms of $\hat\Gamma^\rho_{\mu\nu}$. Then using \eqref{eq:omegacurv} and \eqref{eq:Omega-vs-omega} we find
\begin{equation}
h^{\mu\nu}\check R_{\mu\nu} = -\tilde R^{cd}{}_{cd}(J)=-\mathcal{R}+2(d-1)\nabla_\mu\left(h^{\mu\nu}a_\nu\right)-(d-1)(d-2)h^{\mu\nu}a_\mu a_\nu\,,
\end{equation}
where we used \eqref{eq:TNCb} and $R^{cd}{}_{cd}(J)=\mathcal{R}$ which is merely a definition of $\mathcal{R}$. 

By fully employing the local Schr\"odinger algebra we arrive at the conformally invariant $z=d$ action \cite{Horava:2008ih,Horava:2010zj}
\begin{eqnarray}
S & = & \int d^{d+1}x e\left[A\left(h^{\mu\rho}h^{\nu\sigma}K_{\mu\nu}K_{\rho\sigma}-\frac{1}{d}\left(h^{\mu\nu}K_{\mu\nu}\right)^2\right)\right.\nonumber\\
&&\left. +B\left(\mathcal{R}-2(d-1)\nabla_\mu\left(h^{\mu\nu}a_\nu\right)+(d-1)(d-2)h^{\mu\nu}a_\mu a_\nu\right)^d\right]\,.\label{eq:confinvaction}
\end{eqnarray}
For $z=d$ the dilatation weight of $\tilde\Phi$ is given by $2(d-1)$ so that the terms
\begin{equation}\label{eq:ab}
-a\tilde\Phi\left(\mathcal{R}-2(d-1)\nabla_\mu\left(h^{\mu\nu}a_\nu\right)+(d-1)(d-2)h^{\mu\nu}a_\mu a_\nu\right)+b\tilde\Phi^{\tfrac{d}{d-1}}\,,
\end{equation}
can be added to the action in a conformally invariant manner. Assuming $b\neq 0$ we can integrate out $\tilde\Phi$ which leads to the action \eqref{eq:confinvaction} with a different constant $B$. The case with $b=0$ can be viewed as a constrained system as discussed in section \ref{sec:constraint}. The integrand of \eqref{eq:confinvaction} has been obtained in Lifshitz holography and field theory using different techniques and found to describe the Lifshitz scale anomaly \cite{Griffin:2011xs,Baggio:2011ha,Griffin:2012qx,Christensen:2013rfa,Arav:2014goa} where $A$ and $B$ play the role of two central charges. In \cite{Christensen:2013rfa} it was shown that for $d=z=2$ the integrand of \eqref{eq:confinvaction} together with \eqref{eq:ab} for specific values of $a$ and $b$ arises from the (Scherk--Schwarz) null reduction of the AdS$_5$ conformal anomaly of gravity coupled to an axion.

\section{Discussion \label{sec:discussion} }

We have shown that the dynamics of TTNC geometries, for which there is a hypersurface orthogonal foliation of constant time hypersurfaces, is precisely given by non-projectable Ho\v rava--Lifshitz gravity. The projectable case corresponds to dynamical NC geometries without torsion. One can build a precise dictionary,
  between properties of TNC and HL gravities, which we give below in table \ref{table:dictionary}. 

\begin{table}[h!]
      \centering
      \begin{tabular}{|c|c|}
      \hline
{\bf TNC gravity} & {\bf HL gravity} \\
\hline \hline

twistless torsion: $h^{\mu\rho}h^{\nu\sigma}\left(\partial_\mu\tau_\nu-\partial_\nu\tau_\mu\right)=0$ & non-projectable \\
\hline
no torsion: $\partial_\mu\tau_\nu-\partial_\nu\tau_\mu=0$ & projectable \\
\hline
$\tau_\mu=\psi\partial_\mu\tau$ & scalar khronon $\varphi$ in $u_\mu$ \cite{Blas:2010hb}\\
\hline
$\tau$ invariant under Galilean  & foliation breaks local Lorentz  \\
tangent space group & invariance\\
\hline
torsion vector $a_\mu$ & acceleration $a_\mu$ \cite{Blas:2010hb}\\
\hline
TNC invariant: $-\tau_\mu\tau_\nu+\hat h_{\mu\nu}$ & metric with Lorentz signature $g_{\mu\nu}$\\
\hline
$\tau_i=0$ & ADM decomposition \\
\hline
$\tau_t$ & lapse $N$ \\
\hline
$m_i=-N^{-1}N_i$ & ADM shift vector $N_i$\\
\hline
$\hat h_{ij}$ & metric on constant $t$ slices $\gamma_{ij}$\\
\hline
scalar $\tilde\Phi$ in $m_t=-\frac{1}{2N}\gamma_{ij}N^i N^j+N\tilde\Phi$ & $N^{-1}A$ \cite{Horava:2010zj}\\
\hline
St\"uckelberg scalar $\chi$ & Newtonian prepotential $\nu$ \cite{Horava:2010zj}\\
\hline
Bargmann central extension acting  & local $U(1)$ acting on $A$, $N_i$ and $\nu$\\
on $m_\mu$ and $\chi$ & \\
         \hline
    $\nabla_\mu\hat v^\nu$ & extrinsic curvature \\
    \hline
    two scalar invariants $\nabla_\mu\hat v^\mu\nabla_\nu\hat v^\nu$ and
$\nabla_\nu\hat v^\mu\nabla_\mu\hat v^\nu$ & the $\lambda$ parameter in the kinetic term\\
    allowed by local Galilean symmetries & \\
    \hline
Effective action organized by  & Dimensions: $[N]=L^0$, $[\gamma_{ij}]=L^0$,  \\
Schr\"odinger representations & $[N^i]=L^{1-z}$, $[A]=L^{-2(z-1)}$\\
\hline
Local Schr\"odinger invariance & conformal HL actions (invariant\\
&  under anisotropic Weyl rescalings)\\
\hline
general torsion: no constraint on $\tau_\mu$ & vector khronon \cite{Janiszewski:2012nb}\\
\hline
           \end{tabular}
      \caption{Dictionary between TNC and HL terminology.}\label{table:dictionary}
\end{table}

We conclude with some general comments about interesting future research directions.

TNC geometries have appeared so far as fixed background geometries for non-relativistic field theories and hydrodynamics \cite{Son:2013rqa,Geracie:2014nka,Gromov:2014vla,Moroz:2014ska,Jensen:2014aia,Hartong:2014pma,Jensen:2014ama,Hartong:2015wxa} as well as in holographic setups based on Lifshitz bulk space-times \cite{Christensen:2013lma,Christensen:2013rfa,Hartong:2014oma,Bergshoeff:2014uea,Hartong:2015wxa}. In all these cases the TNC geometry is treated as non-dynamical. This is a valid perspective provided the backreaction onto the geometry can be considered small, e.g. a small amount of energy or mass density should not lead to pathological behavior of the geometry when allowing it to backreact. This renders the question of the consistency of HL gravity in the limit of small fluctuations around flat space-time of crucial importance for applications of TNC geometry to the realm of non-relativistic physics.

In this light we wish to point out that (in the absence of a cosmological constant) the ground state is not Minkowski space-time but flat NC space-time which has different symmetries than Minkowski space-time as worked out in detail in \cite{Hartong:2015wxa}. It would be interesting to work out the properties of perturbations of TTNC gravity around flat NC space-time. In particular we have shown that generically there is no local $U(1)$ symmetry in the problem but that rather one can either integrate out $\tilde\Phi_\chi$ without modifying the effective action in an essential way or in such a way that it imposes a non-trivial constraint on the spatial part of the geometry. It would also be interesting to study the theory from a Hamiltonian perspective and derive the first and second class constraints and compare the resulting counting of degrees of freedom with the linearized analysis. 

Since it is well understood how to couple matter to TNC geometries the question of how to couple matter to HL gravity can be readily addressed in this framework. For example it would be interesting to find Bianchi identities for the TTNC curvature tensor (as studied in section \ref{sec:curvatures}) in such a way that they are compatible with the on-shell diffeomorphism Ward identity for the energy-momentum tensor as defined in  \cite{Hartong:2014oma,Hartong:2014pma,Hartong:2015wxa}. We emphasize once more that matter systems coupled to TNC geometries can have but do not necessarily need to have a particle number symmetry \cite{Hartong:2014oma,Hartong:2015wxa}. It would be important to study what the fate of particle number symmetry is once we make the geometry dynamical. In the matter sector particle number symmetry comes about as a gauge transformation acting on $M_\mu$ in such a manner that the St\"uckelberg scalar $\chi$ can be removed from the matter action \cite{Hartong:2014oma,Hartong:2015wxa} making this formulation consistent with \cite{Jensen:2014aia}. We have seen in section \ref{sec:localU1} that generically the $\chi$ field cannot be removed from the actions describing the dynamics of the TNC geometry. Hence, it seems that the dynamics of the geometry breaks particle number symmetry except when we use the model \eqref{eq:HLaction2} for projectable HL gravity with $\lambda=1$ in which case the central extension of the Bargmann algebra is a true local $U(1)$ symmetry and the $\chi$ field does not appear in the HL action.

Another interesting extension of this work is to consider the case of unconstrained torsion, i.e. TNC gravity, in which case $\tau_\mu$ is no longer restricted to be hypersurface orthogonal. In table \ref{table:dictionary} we refer to this as the vector khronon extension in the last row. The main difference with TTNC geometry is that now the geometry orthogonal to $\tau_\mu$ is no longer torsion free Riemannian geometry but becomes torsionful. This extra torsion is described by an object which we call the twist tensor (see e.g. \cite{Bergshoeff:2014uea})  denoted by $T_{\mu\nu}$ and defined as
\begin{equation}
T_{\mu\nu}=\frac{1}{2}\left(\delta_\mu^\rho+\tau_\mu\hat v^\rho\right)\left(\delta_\nu^\sigma+\tau_\nu\hat v^\sigma\right)\left(\partial_\rho\tau_\sigma-\partial_\sigma\tau_\rho\right)\,.
\end{equation}
Therefore apart from the fact that now the $\tau_\mu$ appearing in the actions of sections \ref{sec:HLactions}--\ref{sec:confHLactions} is no longer of the form $\psi\partial_\mu\tau$ but completely free, we can also add additional terms containing the twist tensor $T_{\mu\nu}$. Another such tensor is $T_{(a)\mu\nu}$ (see again \cite{Bergshoeff:2014uea} where it was denoted by $T_{(b)\mu\nu}$) which is defined as
\begin{equation}
T_{(a)\mu\nu}=\frac{1}{2}\left(\delta_\mu^\rho+\tau_\mu\hat v^\rho\right)\left(\delta_\nu^\sigma+\tau_\nu\hat v^\sigma\right)\left(\partial_\rho a_\sigma-\partial_\sigma a_\rho\right)\,.
\end{equation}
Hence we can add for example a term such as
\begin{equation}
T_{\mu\nu}T_{\rho\sigma}h^{\mu\rho}h^{\nu\sigma}\,,
\end{equation}
which has weight $4-2z$ so that it is relevant for $z>1$. In fact for $z=2$ this term has weight zero and so
one can add an arbitrary function of the twist tensor squared. In the IR however the two-derivative term dominates. 

Another aspect that would be worthwhile examining using our results is whether one could learn more about non-relativistic field theories at finite temperature using holography for HL gravity \cite{Griffin:2011xs,Kiritsis:2012ta,Janiszewski:2012nb,Janiszewski:2012nf,Griffin:2012qx,Wu:2014dha}. Independently of  whether HL gravity is UV complete, assuming it makes sense as a classical theory it may be a useful tool to compute properties such as correlation functions of the (non-relativistic) boundary field theory. In particular, this implies that there must exist  bulk gravity duals to thermal states of the field theory, i.e. classical solutions of HL gravity that resemble black holes as we know them in general relativity. In light of this it would be interesting to re-examine the status of black hole solutions in HL gravity   (see e.g. \cite{Kehagias:2009is,Lu:2009em,Kiritsis:2009rx}). Moreover, it is expected that in a long-wave length regime some version of the fluid/gravity correspondence should exist, enabling the computation of for example transport coefficients in finite temperature non-relativistic field theories on flat (or more generally curved) NC backgrounds.

TNC geometry also appears in the context of WCFTs \cite{Hofman:2014loa} as the geometry to which these $SL(2)\times U(1)$ invariant CFTs couple to. This was called warped geometry and corresponds to TNC geometry in $1+1$  dimensions with $z=\infty$ (or $z=0$ if one interchanges the two coordinates). In that case there is no spatial curvature so the entire dynamics is governed by torsion. It would be interesting to write down the map to the formulation in  \cite{Hofman:2014loa} and furthermore explicitly write the HL actions for that case.

It would also be interesting to explore the relation of TNC gravity to Einstein-aether theory. It was shown in \cite{Jacobson:2010mx} that any solution of Einstein-aether theory with hypersurface orthogonal $\tau_\mu$ is a solution of the IR limit of non-projectable HL gravity. It would thus be natural to expect that any solution of Einstein-aether  theory with unconstrained $\tau_\mu$ is a solution to the IR limit of TNC gravity. In view of the relation \cite{Ambjorn:2010hu,Ambjorn:2013joa} between  causal dynamical triangulations (CDT) and  HL quantum gravity, both involving a global time foliation, there may also be useful applications of TNC geometry in the context of CDT \cite{Ambjorn:2012jv}. Finally, since HL gravity is connected to the mathematics of Ricci flow (see e.g. \cite{Bakas:2010fm}), examining this from the TNC perspective presented in this paper could lead to novel insights.

\section*{Acknowledgments}
\label{ACKNOWL}

We would like to thank  Ioannis Bakas, Jan de Boer, Diego Hofman, Kristan Jensen, Cindy Keeler and Elias Kiritsis for valuable discussions. 
The work of JH is supported by the advanced ERC grant `Symmetries and Dualities in Gravity and M-theory' of Marc Henneaux. The work of NO is supported in part by the Danish National Research Foundation project ``New horizons in particle and condensed matter physics from black holes".  
We thank the Galileo Galilei Institute for Theoretical Physics for the hospitality and the INFN for partial support during the completion of this work.

\appendix

\section{Gauging Poincar\'e}\label{app:gaugingPoincare}

In this appendix we briefly discuss the gauging of the Poincar\'e algebra to show the power of the method in a more familiar context. Consider the Poincar\'e algebra whose generators are $P_a$ and $M_{ab}$ satisfying the commutation relations
\begin{eqnarray}
\left[ M_{ab}\,,P_c \right] & = & \eta_{ac}P_b-\eta_{bc}P_a\,,\\
\left[ M_{ab}\,,M_{cd} \right] & = & \eta_{ac}M_{bd}-\eta_{ad}M_{bc}-\eta_{bc}M_{ad}+\eta_{bd}M_{ac}\,.
\end{eqnarray}
We introduce the Lie algebra valued connection $\mathcal{A}_\mu$ given by
\begin{equation}
\mathcal{A}_\mu=P_a e^a_\mu+\frac{1}{2}M_{ab}\omega_\mu{}^{ab}\,.
\end{equation}
This connection transforms in the adjoint as
\begin{equation}
\delta{\mathcal{A}}_{\mu}=\partial_\mu\Lambda+[{\mathcal{A}}_{\mu}\,,\Lambda]\,,
\end{equation}
where $\Lambda$ is given by
\begin{equation}
\Lambda=P_a\zeta^a+\frac{1}{2}M_{ab}\sigma^{ab}\,.
\end{equation}

What we have done so far is to make the Poincar\'e transformations local. However we would like to connect this to a set of transformations that replace local space-time translations by diffeomorphisms. This can be achieved as follows. We define a new set of local transformations that we denote by $\bar\delta$. The main step is to replace the parameters in $\Lambda$ corresponding to local space-time translations, i.e. $\zeta^a$ by a space-time vector $\xi^\mu$ via $\zeta^a=\xi^\mu e_\mu^a$. This can achieved by the following way of writing $\Lambda$
\begin{equation}
\Lambda=\xi^\mu\mathcal{A}_\mu+\Sigma\,,
\end{equation}
where
\begin{equation}
\Sigma = \frac{1}{2}M_{ab}\lambda^{ab}\,,
\end{equation}
with $\sigma^{ab}=\xi^\mu\omega_\mu{}^{ab}+\lambda^{ab}$. Next we define $\bar\delta\mathcal{A}_\mu$ as
\begin{equation}
\bar\delta\mathcal{A}_\mu=\delta\mathcal{A}_\mu-\xi^\nu\mathcal{F}_{\mu\nu}=\mathcal{L}_\xi\mathcal{A}_\mu+\partial_\mu\Sigma+[{\mathcal{A}}_{\mu}\,,\Sigma]\,,
\end{equation}
where the second equality is an identity and where $\mathcal{F}_{\mu\nu}$ is the curvature
\begin{eqnarray}
\mathcal{F}_{\mu\nu} & = & \partial_\mu\mathcal{A}_\nu-\partial_\nu\mathcal{A}_\mu+[{\mathcal{A}}_{\mu}\,,{\mathcal{A}}_{\nu}]\nonumber\\
&=& P_aR_{\mu\nu}{}^a(P)+\frac{1}{2}M_{ab}R_{\mu\nu}{}^{ab}(M)\,,
\end{eqnarray}
in which we have
\begin{eqnarray}
R_{\mu\nu}{}^a(P) & = & 2\partial_{[\mu} e_{\nu]}^a-2\omega_{[\mu}{}^{ab} e_{\nu]b}\,,\\
R_{\mu\nu}{}^{ab}(M) & = & 2\partial_{[\mu}\omega_{\nu]}{}^{ab}-2\omega_{[\mu}{}^{ca}\omega_{\nu]}{}^b{}_c\,.
\end{eqnarray}
Under the $\bar \delta$ transformations, the connection $e^a_\mu$ associated with the Lorentz momenta $P_a$, transforms as a vielbein while the connection $\omega_\mu{}^{ab}$ associated with the Lorentz boosts $M_{ab}$ become the spin connection coefficients.

In order to define a covariant derivative on the space-time we first introduce a covariant derivative $\mathcal{D}_\mu$ via
\begin{equation}
\mathcal{D}_\mu e_\nu^a=\partial_\mu e_\nu^a-\Gamma^\rho_{\mu\nu}e_\rho^a-\omega_\mu{}^a{}_b e^b_\nu\,,
\end{equation}
which transforms covariantly under the $\bar\delta$ transformations. The affine connection $\Gamma^\rho_{\mu\nu}$ transforms under the $\bar\delta$ transformations as
\begin{equation}
\bar\delta \Gamma^\rho_{\mu\nu}=\partial_\mu\partial_\nu\xi^\rho+\xi^\sigma\partial_\sigma\Gamma^\rho_{\mu\nu}+\Gamma^\rho_{\sigma\nu}\partial_\mu\xi^\sigma+\Gamma^\rho_{\mu\sigma}\partial_\nu\xi^\sigma-\Gamma^\sigma_{\mu\nu}\partial_\sigma\xi^\rho\,,
\end{equation}
so that it is inert under the local Lorentz (tangent space) transformations. We will now relate the properties of the curvatures $R_{\mu\nu}{}^a(P)$ and $R_{\mu\nu}{}^{ab}(M)$ to those of $\Gamma^\rho_{\mu\nu}$. This goes via the vielbein postulate which reads
\begin{equation}
\mathcal{D}_\mu e_\nu^a=\partial_\mu e_\nu^a-\Gamma^\rho_{\mu\nu}e_\rho^a-\omega_\mu{}^a{}_b e^b_\nu=0\,,
\end{equation}
relating $\Gamma^\rho_{\mu\nu}$ to $\omega_\mu{}^{ab}$. Taking the antisymmetric part of the vielbein postulate and moving $\Gamma^\rho_{[\mu\nu]}$ to the other side we obtain
\begin{equation}
R_{\mu\nu}{}^a(P)=2\partial_{[\mu} e_{\nu]}^a-2\omega_{[\mu}{}^{ab} e_{\nu]b}=2\Gamma^\rho_{[\mu\nu]}e_\rho^a\,.
\end{equation}
From this we conclude that the curvature $R_{\mu\nu}{}^a(P)$ is the torsion tensor. To identify the other curvature tensor $R_{\mu\nu}{}^{ab}(M)$ we compute the commutator of two covariant derivatives $\nabla_\mu$ (containing only the connection $\Gamma^\rho_{\mu\nu}$) leading to
\begin{equation}
[\nabla_\mu\,,\nabla_\nu]X_\rho=R_{\mu\nu\rho}{}^\sigma X_\sigma-2\Gamma^\sigma_{[\mu\nu]}\nabla_\sigma X_\rho\,,
\end{equation}
where $R_{\mu\nu\rho}{}^\sigma$ is the Riemann curvature tensor 
\begin{equation}
R_{\mu\nu\sigma}{}^\rho=-\partial_\mu\Gamma^\rho_{\nu\sigma}+\partial_\nu\Gamma^\rho_{\mu\sigma}-\Gamma^\rho_{\mu\lambda}\Gamma^\lambda_{\nu\sigma}+\Gamma^\rho_{\nu\lambda}\Gamma^\lambda_{\mu\sigma}\,,
\end{equation}
that is related to $R_{\mu\nu}{}^{ab}(M)$ (as follows from the vielbein postulate) via
\begin{equation}
R_{\mu\nu\rho}{}^\sigma=-e_{\rho a}e_b^\sigma R_{\mu\nu}{}^{ab}(M)\,,
\end{equation}
so that $R_{\mu\nu}{}^{ab}(M)$ is the Riemann curvature 2-form. The vielbein postulate, because of the fact that $\omega_\mu{}^{ab}$ is antisymmetric in $a$ and $b$, also tells us that the metric $g_{\mu\nu}=\eta_{ab}e^a_\mu e^b_\nu$, which is the unique Lorentz invariant tensor we can build out of the vielbeins, is covariantly constant, i.e. 
\begin{equation}
\nabla_\rho g_{\mu\nu}=0\,.
\end{equation}
As is well known this fixes completely the symmetric part of the connection making it equal to the Levi-Civit\`a connection plus torsion terms which are left unfixed. The common choice in GR to work with torsion-free connections then implies that from the gauging perspective one imposes the curvature constraint $R_{\mu\nu}{}^a(P)=0$. This in turn makes $\omega_\mu{}^{ab}$ a fully dependent connection expressible in terms of the vielbeins and their derivatives. Without fixing the torsion $e_\mu^a$ and $\omega_\mu{}^{ab}$ remain independent.

\renewcommand{\theequation}{\thesection.\arabic{equation}}

%\bibliography{TNCGravity}
%\bibliographystyle{newutphys}

\providecommand{\href}[2]{#2}\begingroup\raggedright\endgroup

\end{document}